\documentclass[12pt, draftclsnofoot, journal, letterpaper, onecolumn]{IEEEtran}
\usepackage{amsfonts}
\usepackage[dvips]{graphicx}
\usepackage{times}
\usepackage{cite}
\usepackage{amsmath}
\usepackage{array}
\usepackage{amssymb}
\newcommand{\mb}{\mathbf}

\usepackage{stfloats}
\usepackage{slashbox}
\usepackage{graphicx}
\usepackage{footnote}
\usepackage{amsthm}
\usepackage{booktabs}
\usepackage{array}
\usepackage{algorithmic}
\usepackage{algorithm}
\usepackage{subeqnarray}
\usepackage{cases}
\usepackage{threeparttable}
\usepackage{color}
\usepackage{hyperref}
\usepackage{epstopdf}
\usepackage{multirow}
\usepackage{tabularx}
\usepackage{enumerate}
\usepackage{multicol}

\newtheorem{corollary}{Corollary}

\newtheorem{lemma}{Lemma}
\newtheorem{remark}{Remark}
\newtheorem{proposition}{Proposition}

\begin{document}
\title{Secure Beamforming for MIMO Two-Way Communications with an Untrusted Relay}
\author{Jianhua~Mo,~\IEEEmembership{Student~Member,~IEEE}, Meixia~Tao,~\IEEEmembership{Senior~Member,~IEEE}, Yuan~Liu,~\IEEEmembership{Member,~IEEE}, and Rui Wang
\thanks{J. Mo is now with Wireless Networking and Communications Group, The University of Texas at Austin. M. Tao, Y. Liu and R. Wang are with the Department of Electronic Engineering, Shanghai Jiao Tong University, P. R. China. (Email: jhmo@utexas.edu, \{mxtao, yuanliu, liouxingrui\}@sjtu.edu.cn).}
\thanks{The material in this paper was partly presented at IEEE Wireless Communications and Networking Conference, Shanghai, China, April 2013\cite{Mo_Jianhua_WCNC13}.}
}
\maketitle

\begin{abstract}
This paper studies the secure beamforming design in a multiple-antenna three-node system where two source nodes exchange messages with the help of an untrusted relay node. The relay acts as both an essential signal forwarder and a potential eavesdropper. Both two-phase and three-phase two-way relay strategies are considered. Our goal is to jointly optimize the source and relay beamformers for maximizing the secrecy sum rate of the two-way communications. We first derive the optimal relay beamformer structures. Then, iterative algorithms are proposed to find source and relay beamformers jointly based on alternating optimization. Furthermore, we conduct asymptotic analysis on the maximum secrecy sum-rate. Our analysis shows that when all transmit powers approach infinity, the two-phase two-way relay scheme achieves the maximum secrecy sum rate if the source beamformers are designed such that the received signals at the relay align in the same direction. This reveals an important advantage of signal alignment technique in against eavesdropping. It is also shown that if the source powers approach zero the three-phase scheme performs the best while the two-phase scheme is even worse than direct transmission. Simulation results have verified the efficiency of the secure beamforming algorithms as well as the analytical findings.
\end{abstract}


\setlength\arraycolsep{2Pt}

\section{Introduction}

\subsection{Motivation}

Cooperative relaying has been shown effective for power reduction, coverage extension and throughput enhancement in wireless communications. Recently, with the advance of wireless information-theoretic security at the physical layer, a new dimension arises for the design of relaying strategies.  In specific, from a perspective of physical-layer security, a relay can be friendly and may help to keep the confidential message from being eavesdropped by others, while an untrusted relay may intentionally eavesdrop the signal when relaying. The case of untrusted relay exists in real life. For example, the relays and sources belong to different network in today's heterogenous network, where the nodes have different security clearances and thus different levels of access to the information. It is therefore important to find out whether the untrusted relay is still beneficial compared with direct transmission and if so what is the new relay strategy.

The goal of this work is to study the physical layer security in two-way relay systems where the relay is untrusted and each node is equipped with multiple antennas. Compared with traditional one-way relaying, the problem in two-way relaying is more interesting. This is because by applying physical layer network coding, the relay only needs to decode the network-coded message rather than each individual message and hence the network coding procedure itself also brings certain security. We will try to address three important questions. First, under what conditions, should we treat the two-way untrusted relay as a passive eavesdropper or seek help from it? This is a challenging problem because different power constraints and antennas configurations may result in different answers. Second, if help is necessary, how to jointly optimize the source and relay beamformers? Typically this would be a non-convex problem and very difficult to solve.
Thirdly, would physical layer network coding, originally known for throughput enhancement in two-way relay systems, bring new insights to the new performance metric of information security?

\subsection{Related Work}


We first briefly review the existing works related to  beamforming design in MIMO two-way relay systems without taking secrecy into account. Then according to the relay being trusted or untrusted, we classify the related work on secure communication in relay systems.

\subsubsection{Beamforming in MIMO Two-way Relay Systems}
When the source nodes are each equipped with a single antenna, \cite{Zhang_Rui_JSAC09} proposed the optimal relay beamforming structure and a convex optimization algorithm to find the capacity region. For the case of multi-antenna uses, the problem is much more difficult. The work \cite{Xu_Shengyang_TWC11} showed an optimal structure of the relay precoding matrix and proposed an alternating optimization method (optimize the relay precoding matrix and source precoding matrices alternatively) to maximize the achievable weighted sum rate. Based on another criterion of mean-square-errer (MSE), \cite{Wang_Rui_TSP12} proposed an iterative method for the joint source and relay precoding design.

\subsubsection{Trusted Relay}
In this case, the relay is a legitimate user or acts like a legitimate user who will help to counter external eavesdroppers and increase the security of the networks. Most of the work has focused on traditional one-way relaying secret communication (e.g., \cite{Lai_IT08,Jeong_TSP11,Dong_TSP10,Krikidis_TWC09,Ng_TWC11,Huang_TSP11,Mo_Jianhua_CL12}).

Only a few attempts have been made very recently to study two-way relaying secret communication \cite{Chen_Jingchao_TIFS12,Ding_Zhiguo_TVT12,Mukherjee_SPAWC10,Wang_Huiming_TSP12,Shimizu_TIFS11}. Specifically, \cite{Chen_Jingchao_TIFS12} and \cite{Ding_Zhiguo_TVT12} investigated the relay and jammer selection problem in the two-way relay networks. The authors in \cite{Mukherjee_SPAWC10} studied beamforming design in MIMO two-way relaying for maximizing secrecy sum rate which is proven to be achievable in \cite{Tekin_IT08}. Joint distributed beamforming and power allocation was considered in \cite{Wang_Huiming_TSP12} for maximizing secrecy sum rate in two-way relaying networks with multiple single-antenna relays. Several secret key agreement schemes were proposed in \cite{Shimizu_TIFS11}.

\subsubsection{Untrusted Relay}
Untrusted relay channels with confidential messages was first studied in \cite{Oohama_ITW01}, where an achievable secrecy rate was obtained. A destination-based jamming (DBJ) technique was proposed in \cite{He_Xiang_GC08, He_Xiang_JWCN09} without source-destination link. The performance of DBJ in fading channel and multi-relay scenarios was analyzed in \cite{Sun_Li_TVT12}. When the source-destination link exists, authors in \cite{He_Xiang_IT10} discussed whether cooperating with the untrusted relay is better than treating it as a passive eavesdropper.
A Stackelberg game between the two sources and the external friendly jammers in a two-way relay system was formulated as a power control scheme in \cite{Zhang_Rongqing_TVT12}.
In \cite{Jeong_TSP12}, the authors considered MIMO one-way amplify-and-forward (AF) relay systems and jointly deigned the source and relay beamforming using alternating optimization.
\cite{Huang_Jing_TSP13} examines the secrecy outage probability in one-way non-regenerative relay systems.

From these existing literature, it is found that the problem of joint source and relay beamforming for MIMO two-way untrusted relaying has not been considered yet.

\subsection{Contribution}
In this paper, we investigate physical layer security in MIMO two-way relay systems, where the two sources exchange confidential information with each other through an untrusted relay. The relay acts as both an essential helper and a potential eavesdropper, but does not make any malicious attack. {\color{black}In our previous work \cite{Mo_Jianhua_WCNC13}, we considered the two-phase scheme. In this extension, we study both two-phase and three-phase two-way relay schemes.} In particular, we formulate the joint secure source and relay beamforming design for each two-way relay scheme. The objective is to maximize the secrecy sum rate of the bidirectional links subject to the source and relay power constraints. Furthermore, we conduct asymptotic analysis on maximum secrecy sum rate of the different two-way relay schemes in comparison with direct transmission.

The main contributions and results of this paper are summarized as follows:
\begin{itemize}
  \item The optimal structure of the relay beamforming matrix for fixed source beamformers is derived. With this structure, the number of unknowns in the relay beamformer is significantly reduced and thus the joint source and relay beamformer design can be simplified.

  \item Iterative algorithms based on alternating optimization are proposed to find a solution of the joint source and relay beamformers. These algorithms are convergent but cannot ensure global optimality due to the nonconvexity of the optimization problems.

  \item Via asymptotic analysis, we show that when the powers of the source and relay nodes approach infinity, the two-phase scheme achieves the maximum secrecy rate if the transceiver beamformers are designed such that the received signals at the relay align in the same direction. This reveals an important advantage of \emph{ signal alignment} techniques in against eavesdropping. It
      gives a new perspective to achieve the physical layer security, and also lowers the source antenna number requirement for ensuring security.

  \item It is also shown via asymptotic analysis that when the power of the relay goes to infinity and that of the two sources approach zeros, the three-phase two-way relay scheme performs the best while the two-phase performs even worse than direct transmission.
\end{itemize}

\subsection{Organization and Notations}

The rest of the paper is organized as follows. Section II describes the system model and problem formulations. The optimal secure beamformers for two- and three-phase two-way relay schemes are presented in Section III. Asymptotical results are detailed in Section IV. Comprehensive simulation results are given in Section V. Finally, we conclude this paper in Section VI.

\textsl{Notations}: Scalars, vectors and matrices are denoted by lower-case, lower-case bold-face and upper-case bold-face letters, respectively. $[x]^+$ denotes $\max\left(0,x\right)$. $\mathrm{Tr}(\mathbf{A})$, $\mathbf{A}^{-1}$, $\mathrm{Rank}(\mathbf{A})$, $\|\mathbf{A}\|_F$, $\mathbf{A}^*$ and $\mathbf{A}^H$ denote the trace, inverse, rank, Frobenius norm, conjugate and Hermite of matrix $\mathbf{A}$, respectively. {\color{black} ${\rm span}(\mb{A})$ represents the column space (range space) of $\mb{A}$ and ${\rm dim}(\mb{A})$ denotes the dimension of $\mb{A}$. The projection matrix onto the null space of $\mb{A}$ is denoted by $\mb{A}^{\mathcal{N}}$.} $\|\mathbf{q}\|$ denotes the norm of the vector $\mathbf{q}$. $\sigma_{\max}(\mathbf{A})$ is the largest singular value of $\mathbf{A}$. $\lambda_{\max}({\mathbf{A}})$ is the largest eigenvalue of the matrix $\mathbf{A}$ and $\pmb{\psi}_{\max}(\mathbf{A})$ is the eigenvector of $\mathbf{A}$ corresponding to the largest eigenvalue. $\lambda_{\max}(\mathbf{A}, \mathbf{B})$ is the largest generalized eigenvalue of the matrices $\mathbf{A}$ and $\mathbf{B}$. $\pmb{\psi}_{\max}(\mathbf{A}, \mathbf{B})$ is the generalized eigenvector of $(\mathbf{A},\mathbf{B})$ corresponding to the largest generalized eigenvalue.
We use $P_i^{DT}$, $P_i^{2P}$ and $P_i^{3P}$ to represent the transmit power of node $i \in \{A, B, R\}$ in two-way direct transmission, two-phase two-way relaying and three-phase two-way relaying, respectively. Throughout this paper, $\mathbf{n}_i$ denotes the zero mean circularly symmetric complex Gaussian noise vector at node $i \in \{A, B, R\}$ with $\mathbf{n}_i \sim \mathcal{CN}\left(\mathbf{0}, \mathbf{I}\right)$.

\section{System Model and Problem Formulation}

\begin{figure}[t]
\begin{centering}
\includegraphics[scale=.80]{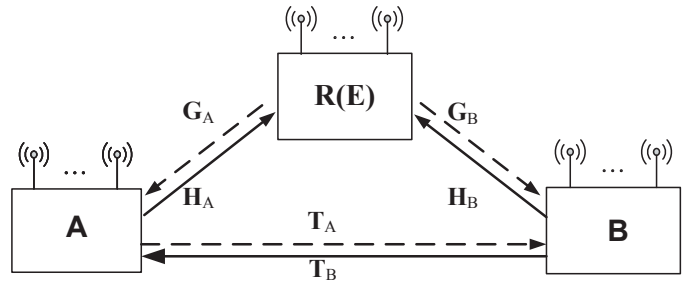}
\vspace{-0.1cm}
\centering
 \caption{MIMO two-way relay model.}\label{fig:system}
\end{centering}
\vspace{-0.3cm}
\end{figure}

We consider a two-way relay system as shown in Fig. \ref{fig:system}, where two source nodes $A$ and $B$ exchange information with each other with the assistance of a relay node $R$. The relay acts as both an essential helper and a potential eavesdropper but does not make any malicious attack. Note that the decode-and-forward (DF) relay strategy is not applicable here since the relay is untrusted and not expected to decode the received signal from the source nodes. As such, we assume the relay adopts AF strategy, which also has low complexity.
The number of antennas at nodes $A$, $B$ and $R$ are denoted as $N_A$, $N_B$ and $N_R$, respectively. As shown in Fig. \ref{fig:system}, $\mathbf{T}_A \in \mathbb{C}^{N_B\times N_A}$, $\mathbf{T}_B \in \mathbb{C}^{N_A\times N_B}$, $\mathbf{H}_A \in \mathbb{C}^{N_R\times N_A}$, $\mathbf{G}_A \in \mathbb{C}^{N_A\times N_R}$, $\mathbf{H}_B \in \mathbb{C}^{N_R\times N_B}$, $\mathbf{G}_B \in \mathbb{C}^{N_B\times N_R}$ denote the channel matrices of link $A\to B$, $B\to A$, $A\to R$, $R\to A$, $B\to R$ and $R\to B$, respectively. If the system operates in time division duplex (TDD) mode and channel reciprocity holds, then we have $\mathbf{T}_A=\mathbf{T}_B^T$, $\mathbf{H}_A=\mathbf{G}_A^T$, and $\mathbf{H}_B=\mathbf{G}_B^T$. For simplicity, we only consider single data stream for each source node in this paper. Denote the transmitted symbol at the source $i$ as $s_i \in \mathbb{C}$ with $\mathbb{E}(|s_i|^2)=1$, and the associated beamforming vector as $\mathbf{q}_i\in \mathbb{C}^{N_i\times 1}$, for $i\in\{A, B\}$.

Different two-way relay schemes have been studied in the literature \cite{Rankov_JSAC07,Kim}. In this paper, we study the two-phase and three-phase two-way relay schemes. For the purpose of comparison, the two-way direct transmission is also considered in Appendix \ref{app_DT}, wherein the relay node is treated as a pure eavesdropper \cite{Jeong_TSP12,Khisti_IT10a}.

\subsection{Two-Phase Two-Way Relay Scheme}
In the first phase of the two-phase two-way relay scheme, $A$ and $B$ simultaneously transmit signals to the relay node $R$. The received signal at relay is,
\begin{eqnarray}\label{2_phase_MAC}
{\mathbf{y}_R^{2P}} &=& {\mathbf{H}_A}{\mathbf{q}_A}{s_A} + {\mathbf{H}_B}{\mathbf{q}_B}{s_B} + {\mathbf{n}_R}.
\end{eqnarray}

In the second phase, the relay node amplifies $\mathbf{y}_R^{2P}$ by multiplying it with a precoding matrix $\mathbf{F}$ and broadcasts it to both A and B.
The transmit signal vector from the relay node is expressed as
\begin{equation}\label{2_phase_Relay_Beamforming}
  {\mathbf{x}_R^{2P}} = \mathbf{F}{\mathbf{y}_R^{2P}}.
\end{equation}

After subtracting the back propagated self-interference, each source node $i$ obtains the equivalent received signals,
\begin{equation}\label{2_phase_Broadcast_equivalent}
  \mathbf{y}_i^{2P} = \mathbf{G}_i \mathbf{F} \mathbf{H}_{\bar{i}} \mathbf{q}_{\bar{i}} s_{\bar{i}} + \mathbf{G}_i \mathbf{F} \mathbf{n}_R + \mathbf{n}_i, ~ i\in \{A, B\},
\end{equation}
where $\bar{i} = \{A, B\}\setminus i$.

The information rate from node $i$ to node $\bar{i}$ is
\begin{equation}
{\mathcal{R}_{i\bar{i}}^{2P}} = \frac{1}{2} \log_2 \left( {1 + \mathbf{q}_i^H \mathbf{H}_i^H{\mathbf{F}^H} \mathbf{G}_{\bar{i}}^H \mathbf{K}_{\bar{i}}^{-1} {\mathbf{G}_{\bar{i}}} \mathbf{F}{\mathbf{H}_i} {\mathbf{q}_i}} \right), \label{eq_2P_R_AB}
\end{equation}
where
\begin{equation}
  {\mathbf{K}_{\bar{i}}} =  {\mathbf{G}_{\bar{i}}}\mathbf{F}{\mathbf{F}^H}\mathbf{G}_{\bar{i}}^H + \mathbf{I}.
\end{equation}

If the untrusted relay wants to eavesdrop the signals from both source nodes, it may try to fully decode the two messages $s_A$ and $s_B$. Therefore, the achievable information rate at the untrusted relay can be expressed as the maximum sum-rate of a two-user MIMO multiple-access channel, given by
\begin{eqnarray}
 {\mathcal{R}_R^{2P}} &\triangleq &  \textit{I}\left(\mathbf{y}_{R}^{2P}; s_A, s_B \right) \nonumber \\
 &=& \frac{1}{2} \log_2 \left| \mathbf{I} + \left[ {\begin{array}{*{20}{c}}
{{\mathbf{H}_A}{\mathbf{q}_A}}&{{\mathbf{H}_B}{\mathbf{q}_B}}
\end{array}} \right] \left[ {\begin{array}{*{20}{c}}
{{\mathbf{q}_A^H}{\mathbf{H}_A^H}}\\
{{\mathbf{q}_B^H}{\mathbf{H}_B^H}}
\end{array}} \right] \right| \nonumber\\
&\stackrel{(a)}=& \frac{1}{2} \log_2 \left| \mathbf{I} + \left[ {\begin{array}{*{20}{c}}
{{\mathbf{q}_A^H}{\mathbf{H}_A^H}}\\
{{\mathbf{q}_B^H}{\mathbf{H}_B^H}}
\end{array}} \right]  \left[ {\begin{array}{*{20}{c}}
{{\mathbf{H}_A}{\mathbf{q}_A}}&{{\mathbf{H}_B}{\mathbf{q}_B}}
\end{array}} \right] \right| \nonumber\\
&=& \frac{1}{2} \log_2 \left|  {\begin{array}{*{20}{c}}
{1+{\mathbf{q}_A^H}{\mathbf{H}_A^H}{\mathbf{H}_A}{\mathbf{q}_A}} & {{\mathbf{q}_A^H}{\mathbf{H}_A^H}{\mathbf{H}_B}{\mathbf{q}_B}} \\
{\mathbf{q}_B^H}{\mathbf{H}_B^H}{\mathbf{H}_A}{\mathbf{q}_A} & {1+{\mathbf{q}_B^H}{\mathbf{H}_B^H}{\mathbf{H}_B}{\mathbf{q}_B}}
\end{array}}     \right| \nonumber\\
 &=& \frac{1}{2} \log_2 \Big(1+ \|\mathbf{H}_A \mathbf{q}_A \|^2 + \|\mathbf{H}_B \mathbf{q}_B \|^2 \nonumber \\
 & & + \|\mathbf{H}_A \mathbf{q}_A \|^2 \|\mathbf{H}_B \mathbf{q}_B \|^2 - \|\mathbf{q}_B^H \mathbf{H}_B^H  \mathbf{H}_A \mathbf{q}_A \|^2    \Big) \label{eq_2P_AB_R}.
\end{eqnarray}
where $(a)$ is from the identity $\left|\mathbf{I}+\mathbf{A}\mathbf{B}\right|=\left|\mathbf{I}+\mathbf{B}
\mathbf{A}\right|$.

The achievable secrecy sum rate\cite{Tekin_IT08} of the two source nodes is thus given by,
\begin{equation}\label{eq_2P_R_s}
   \mathcal{R}_{s}^{2P} = \left[\mathcal{R}_{AB}^{2P} + \mathcal{R}_{BA}^{2P} - {\mathcal{R}_R^{2P}} \right]^{+}
\end{equation}

Our goal is to maximize the secrecy sum rate by jointly optimizing the relay and source beamformers $\mathbf{F}$, $\mathbf{q}_A$ and $\mathbf{q}_B$. The problem can be formulated as
\begin{subequations}\label{eq_2P_problem}
\begin{align}
  & & &\mathop{\max}\limits_{\{\mathbf{F},{\mathbf{q}_A},{\mathbf{q}_B}\}} \mathcal{R}_s^{2P} \\
  &\text{s. t.} & &\|\mathbf{q}_i\|^2 \leq P_i^{2P}, ~~i \in \{A,B\},  \\
  & & &\mathrm{Tr} \big( \mathbf{F} \mathbf{H}_A \mathbf{q}_A \mathbf{q}_A^H \mathbf{H}_A^H \mathbf{F}^H + \mathbf{F} \mathbf{H}_B \mathbf{q}_B \mathbf{q}_B^H \mathbf{H}_B^H \mathbf{F}^H \nonumber \\
  & & &+ \mathbf{F} \mathbf{F}^H　\big) \leq P_R^{2P}. \label{eq_2P_power_cons}
  \end{align}
\end{subequations}
where \eqref{eq_2P_power_cons} is the relay power constraint.

\subsection{Three-Phase Two-Way Relay Scheme}
In the first phase of the three-phase two-way relay scheme, source node $A$ transmits.
The received signals at the relay $R$ and the source node $B$ are respectively given by
\begin{eqnarray}
    {{\mathbf{y}}_{R1}^{3P}} &=& {{\mathbf{H}}_A}{{\mathbf{q}}_A}{s_A} + {{\mathbf{n}}_{R1}}, \label{eq_3P_A_R}\\
    {{\mathbf{y}}_{B1}^{3P}} &=& {{\mathbf{T}}_A}{{\mathbf{q}}_A}{s_A} + {{\mathbf{n}}_{B1}}.
    \label{eq_3P_A_B}
\end{eqnarray}
where $\mathbf{n}_{R1}$ and $\mathbf{n}_{B1}$ are the noises at the relay node and source node $B$ in the first phase, respectively.

In the second phase, node $B$ transmits. The received signals at the relay and the source node $A$ are respectively given by
\begin{eqnarray}
    {{\mathbf{y}}_{R2}^{3P}} &=& {{\mathbf{H}}_B}{{\mathbf{q}}_B}{s_B} + {{\mathbf{n}}_{R2}}, \label{eq_3P_B_R}\\
    {{\mathbf{y}}_{A1}^{3P}} &=& {{\mathbf{T}}_B}{{\mathbf{q}}_B}{s_B} + {{\mathbf{n}}_{A1}}. \label{eq_3P_B_A}
\end{eqnarray}
where $\mathbf{n}_{R2}$ and $\mathbf{n}_{A1}$ are the noises at the relay node and source node $A$ in the second phase, respectively.

In the third phase, the relay node $R$ amplifies the received signals $\mathbf{y}_{R1}^{3P}$ and $\mathbf{y}_{R2}^{3P}$ by multiplying them with $\mathbf{F}_A$ and $\mathbf{F}_B$, respectively. The broadcast signal from the relay is thus
\begin{equation}
  \mathbf{x}_R^{3P} = \mathbf{F}_A \mathbf{y}_{R1}^{3P} + \mathbf{F}_B \mathbf{y}_{R2}^{3P}.
\end{equation}
After subtracting the self-interference, the two source nodes obtain the signals as
%
\begin{eqnarray}\label{eq_3P_BC_equivalent}
  \mathbf{y}_{A2}^{3P} = {{\mathbf{G}}_A} {{\mathbf{F}}_B} {{\mathbf{H}}_B}{{\mathbf{q}}_B}{s_B} + {{\mathbf{G}}_A}\left( {{{\mathbf{F}}_A}{{\mathbf{n}}_{R1}} + {{\mathbf{F}}_B}{{\mathbf{n}}_{R2}}} \right) + {{\mathbf{n}}_{A2}}, \label{eq_3P_BC_equivalent_A} \\
  \mathbf{y}_{B2}^{3P} = {{\mathbf{G}}_B} {{\mathbf{F}}_A} {{\mathbf{H}}_A}{{\mathbf{q}}_A}{s_A} + {{\mathbf{G}}_B}\left( {{{\mathbf{F}}_A}{{\mathbf{n}}_{R1}} + {{\mathbf{F}}_B}{{\mathbf{n}}_{R2}}} \right) + {{\mathbf{n}}_{B2}}. \label{eq_3P_BC_equivalent_B}
\end{eqnarray}

Combining \eqref{eq_3P_B_A} and \eqref{eq_3P_BC_equivalent_A}, we obtain the information rate from $B$ to $A$ as,
\begin{eqnarray}\label{eq_3P_B_A_rate}
  \mathcal{R}_{BA}^{3P} &\triangleq& \textit{I}(\mathbf{y}_{A1}^{3P}, \mathbf{y}_{A2}^{3P} ; \mathbf{s}_B) \\
  &=& \frac{1}{3} \log_2 \Big( 1 + {\mathbf{q}}_B^H{\mathbf{T}}_B^H{{\mathbf{T}}_B}{{\mathbf{q}}_B}
   + {\mathbf{q}}_B^H{\mathbf{H}}_B^H{\mathbf{F}}_B^H{\mathbf{G}}_A^H \cdot \nonumber \\
   & &{{\left( {{{\mathbf{G}}_A}\left( {{{\mathbf{F}}_A}{\mathbf{F}}_A^H + {{\mathbf{F}}_B}{\mathbf{F}}_B^H} \right){\mathbf{G}}_A^H + {\mathbf{I}}} \right)}^{ - 1}}{{\mathbf{G}}_A}{{\mathbf{F}}_B}{{\mathbf{H}}_B}{{\mathbf{q}}_B}\Big). \nonumber
\end{eqnarray}
Likewise, the information rate from $A$ to $B$ is,
\begin{eqnarray}\label{eq_3P_A_B_rate}
  \mathcal{R}_{AB}^{3P} &=& \frac{1}{3} \log_2 \Big( 1 + {\mathbf{q}}_A^H{\mathbf{T}}_A^H{{\mathbf{T}}_A}{{\mathbf{q}}_A} +
  {\mathbf{q}}_A^H{\mathbf{H}}_A^H{\mathbf{F}}_A^H{\mathbf{G}}_B^H \cdot \nonumber \\
  & &{{\left( {{{\mathbf{G}}_B}\left( {{{\mathbf{F}}_A}{\mathbf{F}}_A^H + {{\mathbf{F}}_B}{\mathbf{F}}_B^H} \right){\mathbf{G}}_B^H + {\mathbf{I}}} \right)}^{ - 1}}{{\mathbf{G}}_B}{{\mathbf{F}}_A}{{\mathbf{H}}_A}{{\mathbf{q}}_A}\Big). \nonumber
\end{eqnarray}

The information sum rate leaked to the untrusted relay can be obtained from \eqref{eq_3P_A_R} and \eqref{eq_3P_B_R}:
\begin{eqnarray}
  \mathcal{R}_{R}^{3P} &\triangleq& \textit{I} (\mathbf{y}_{R1}^{3P} ; \mathbf{s}_A) + \textit{I} (\mathbf{y}_{R2}^{3P} ;\mathbf{s}_B )  \\
  &=& \frac{1}{3} \log_2 {\left(\left( {1 + {\mathbf{q}}_A^H{\mathbf{H}}_A^H{{\mathbf{H}}_A}{{\mathbf{q}}_A}} \right)\left( {1 + {\mathbf{q}}_B^H{\mathbf{H}}_B^H{{\mathbf{H}}_B}{{\mathbf{q}}_B}} \right)\right)} \nonumber
\end{eqnarray}

Thus, the secrecy sum rate is given by
\begin{equation}\label{eq_3P_secrecy_sum_rate}
  \mathcal{R}_s^{3P} = [\mathcal{R}_{BA}^{3P}+\mathcal{R}_{AB}^{3P}-\mathcal{R}_R^{3P}]^+.
\end{equation}

We  can formulate the secrecy sum rate maximization problem for three-phase two-way relay scheme as
\begin{subequations}\label{eq_3P_problem}
\begin{align}
   & & & \mathop{\max}\limits_{\{\mathbf{F}_A,\mathbf{F}_B,{\mathbf{q}_A},{\mathbf{q}_B}\}} \mathcal{R}_s^{3P} \\
   &\mbox{s. t.} & & \|\mathbf{q}_i\|^2 \leq P_i^{3P}, ~~i \in \{A,B\}, \\
   & & & \mathrm{Tr}\Big(\mathbf{F}_A \mathbf{H}_A \mathbf{q}_A \mathbf{q}_A^H \mathbf{H}_A^H \mathbf{F}_A^H + \mathbf{F}_B \mathbf{H}_B \mathbf{q}_B \mathbf{q}_B^H \mathbf{H}_B^H \mathbf{F}_B^H  \nonumber \\
   & & &+ \mathbf{F}_A \mathbf{F}_A^H + \mathbf{F}_B \mathbf{F}_B^H　\Big) \leq P_R^{3P},
   \label{eq_3P_power_cons}
\end{align}
\end{subequations}
where \eqref{eq_3P_power_cons} is the relay power constraint.

In these two schemes, we assume that one of the source nodes, say $A$ is responsible for the joint design of source and relay beamformers. After finishing the design, $A$ sends the corresponding designed beamformer to $B$ and the relay. Then, the two source nodes and the untrusted relay will use their beamformers to process the transmit signals.

\section{Secure Beamfomring Designs}
 After introducing the problem formulations in \eqref{eq_2P_problem} and \eqref{eq_3P_problem} for the two-phase and three-phase two-way relay schemes, respectively, we now present algorithms to design these secure beamformers in this section.

\subsection{Secure Beamforming in Two-Phase Two-Way Relay Scheme}

We first obtain the optimal structure of the secure relay beamforming matrix $\mathbf{F}$. Then we present an iterative algorithm to find a local optimal solution for the joint secure source and relay beamformers. Define the following two QR decompositions:
\begin{equation}
  [\mathbf{G}_A^H ~~ \mathbf{G}_B^H ] = \mathbf{V} \mathbf{R}_{1}^{2P},
\end{equation}
\begin{equation}\label{QR_2_phase_Hq}
  [\mathbf{H}_A \mathbf{q}_A ~~ \mathbf{H}_B \mathbf{q}_B] = \mathbf{U} \mathbf{R}_{2}^{2P}.
\end{equation}
where $\mathbf{V} \in \mathbb{C}^{N_R \times \min \{N_A + N_B, N_R\}}$, $\mathbf{U} \in \mathbb{C}^{N_R \times 2}$ are orthonormal matrices and $\mathbf{R}_1^{2P}$, $\mathbf{R}_2^{2P}$ are upper triangle matrices.

\begin{lemma}\label{lemma_2P_structure}
  In the two-phase two-way relay scheme, the optimal relay beamforming matrix $\mathbf{F} \in \mathbb{C}^{N_R \times N_R}$ that maximizes the secrecy sum rate has the following structure:
  \begin{equation}\label{eq_2P_structure}
    \mathbf{F}=\mathbf{V}\mathbf{A}\mathbf{U}^H,
  \end{equation}
  where $\mathbf{A} \in \mathbb{C}^{ \min\{N_A+N_B, N_R\} \times 2}$ is an unknown matrix.
\end{lemma}
\begin{IEEEproof}
Note that the relay beamforming matrix $\mathbf{F}$ only influences the information rate $\mathcal{R}_{AB}^{2P}$ and $\mathcal{R}_{BA}^{2P}$. Therefore, the optimal $\mathbf{F}$ that maximize the secrecy sum rate is actually the same to the $\mathbf{F}$ that maximizes the information sum rate $\mathcal{R}_{AB}^{2P}+\mathcal{R}_{BA}^{2P}$. Due to the rank-one precoding at each source node, we have the equivalent channel $\mathbf{H}_i \mathbf{q}_i$ from source node $i$ to relay. Therefore, applying the results in \cite{Xu_Shengyang_TWC11}, we readily have Lemma \ref{lemma_2P_structure}.
\end{IEEEproof}

According to Lemma \ref{lemma_2P_structure}, the number of unknowns in $\mathbf{F}$ is reduced from $N_R^2$ to $2\min\{N_R, N_A+N_B\}$, which reduces the computational complexity of the joint source and relay beamforming design.

We note that it is not easy to find the optimal solution to the problem \eqref{eq_2P_problem}. Even after substituting the optimal structure of $\mathbf{F}$ \eqref{eq_2P_structure} into \eqref{eq_2P_R_s}, the problem is still nonconvex since the secrecy sum rate is not a convex function of $\mathbf{q}_A$, $\mathbf{q}_B$ and $\mathbf{A}$. Therefore, we optimize the source beamforming vectors $\mathbf{q}_A$, $\mathbf{q_B}$ and the unknown matrix $\mathbf{A}$ in the relay beamforming matrix $\mathbf{F}$ in an alternating manner. Given $\mathbf{q}_A$ and $\mathbf{q}_B$, we use the gradient method shown in Appendix \ref{app_2P_subproblem_A} to search $\mathbf{A}$. Given $\mathbf{F}$ and $\mathbf{q}_i$, we can find the optimal $\mathbf{q}_{\bar{i}}$, where the optimization method is shown in Appendix \ref{app_2P_q_B_design}. Formally, we present the method in Algorithm 1 as follows. Here, the initial points of the complex vectors $\mathbf{q}_A$ and $\mathbf{q}_B$ can be randomly generated as long as they satisfy the given power constraint.

\begin{algorithm}
\caption{Iterative algorithm for secure beamforming in two-phase two-way relay scheme}
\begin{algorithmic}[1]
\STATE \textbf{Initialize} $\mathbf{A}$, $\mathbf{q}_A$ and $\mathbf{q}_B$.
\STATE \textbf{Repeat}
    \begin{enumerate}[(a)]
  \item Optimize $\mathbf{A}$ given $\mathbf{q}_A$ and $\mathbf{q}_B$ based on gradient method given in Appendix \ref{app_2P_subproblem_A};
  \item Optimize $\mathbf{q}_B$ given $\mathbf{A}$ and $\mathbf{q}_A$ according to Appendix \ref{app_2P_q_B_design};
  \item Optimize $\mathbf{q}_A$ given $\mathbf{A}$ and $\mathbf{q}_B$ according to Appendix \ref{app_2P_q_B_design} by swapping $A$ and $B$;
    \end{enumerate}
\STATE \textbf{Until} the secrecy sum rate does not increase.
 \end{algorithmic}
\end{algorithm}

Note that the algorithm always converges because the secrecy sum rate is finite and does not decrease in every iteration.

\subsection{Secure Beamforming in Three-Phase Two-Way Relay Scheme}

Similar to the two-phase case, we define the following QR decomposition:
\begin{equation}\label{eq_qr_3P}
  [\mathbf{G}_A^H ~~ \mathbf{G}_B^H ] = \mathbf{V} \mathbf{R}^{3P},
\end{equation}
where $\mathbf{V} \in \mathbb{C}^{N_R \times \min \{N_A + N_B, N_R\}}$ is an orthonormal matrix and $\mathbf{R}^{3P} \in
\mathbb{C}^{\min \{N_A+N_B, N_R\}\times (N_A+N_B)}$ is an upper triangle matrix. Then we give the optimal structure of the relay beamforming matrices $\mathbf{F}_A$ and $\mathbf{F}_B$ in the following lemma.
\begin{lemma}
\label{lemma_3P_structure}
  In the three-phase two-way relay scheme, the optimal relay beamforming matrices $\mathbf{F}_A$, $\mathbf{F}_B$ that maximize the secrecy sum rate have the following structure:
  \begin{eqnarray}\label{eq_3P_structure}
    \mathbf{F}_A=\mathbf{V}\mathbf{a}_1 \frac{\left(\mathbf{H}_A \mathbf{q}_A \right)^H}{\|\mathbf{H}_A \mathbf{q}_A\|}, ~~
    \mathbf{F}_B=\mathbf{V}\mathbf{a}_2 \frac{\left(\mathbf{H}_B \mathbf{q}_B \right)^H}{\|\mathbf{H}_B \mathbf{q}_B\|},
  \end{eqnarray}
  where $\mathbf{a}_1 \in \mathbb{C}^{\min \{N_A+N_B, N_R\} \times 1}$, $\mathbf{a}_2 \in \mathbb{C}^{ \min \{N_A+N_B, N_R\} \times 1}$ are unknown vectors.
\end{lemma}
\begin{IEEEproof} See Appendix \ref{app_3P_structure}.
\end{IEEEproof}

Lemma \ref{lemma_3P_structure} simplifies the design of two beamforming matrices $\mb{F}_i$  to the design of two beamforming vectors $\mb{a}_i$. Thus, the number of unknowns is reduced to $2\min\{N_R, N_A+N_B\}$. Note that the number of unknowns in the relay beamforming matrices is the same for two- and three-phase schemes.

{\color{black}Lemma \ref{lemma_2P_structure} and \ref{lemma_3P_structure} show that the optimal relay beamforming contains three parts: (i) matching to the received signal; (ii) combination or other operation of the information-bearing signals; (iii) beamforming to the intended receiver. This structure is similar to the optimal relay beamforming structure in other systems, for example, the two-way relaying system without secrecy constraint in \cite{Zhang_Rui_JSAC09, Xu_Shengyang_TWC11} and one-way relaying system with secrecy constraint in \cite{Jeong_TSP12}. These structure are also used in our following asymptotical analysis.}

Since problem \eqref{eq_3P_problem} is also nonconvex, we adopt the iterative method to obtain a solution where $\mathbf{q}_A$, $\mathbf{q}_B$, $\mathbf{a}_1$ and $\mathbf{a}_2$ are alternatively optimized until the secrecy sum rate does not increase. The algorithm, denoted as Algorithm 2, is very similar to the Algorithm 1 and omitted.

Since the problems \eqref{eq_2P_problem} and \eqref{eq_3P_problem} are both nonconvex, the iterative algorithms presented in the previous section cannot ensure global optimality. However, letting the transmit power on each node approach zero or infinity, we can derive interesting intuitions which lead  to the asymptotically optimal solution for secure beamforming. In the next section, we present such asymptotic analysis.

\section{Asymptotic Analysis}
 The goal of this section is to find the asymptotical optimal secure beamforming design when the relay power $P_R$ approaches infinity. We first present the analysis when the two source powers are also infinite in Subsection \ref{subsection_P_i_infty}, followed by the analysis when the two source powers approach zero in Subsection \ref{subsection_P_i_0}. Finally, we briefly discuss the case where the relay power $P_R$ approaches zero. For comparison purpose, the asymptotic result for the direct transmission is presented in this section as well.

\subsection{The Case of High Relay and Source Powers}
\label{subsection_P_i_infty}
\begin{proposition}[2P]
\label{proposition_2P_P_R_infty_P_i_infty}
    When $P_R \to \infty$, $P_A \to \infty$ and $P_B \to \infty$,
  the maximum secrecy sum rate of the two-phase two-way relay scheme is,
\begin{eqnarray}\label{eq_2P_angle_bound}
   & &  \mathcal{R}_{\max}^{2P} \nonumber \\
    &\approx &\left\{
\begin{split}
   & \mathop {\max }\limits_{(\mb{q}_A, \mb{q}_B) \in \mathcal{S}}
   \frac{1}{2}{\log _2}\frac{{{{\left\| {{{\mb{H}}_A}{{\mb{q}}_A}} \right\|}^2}{{\left\| {{{\mb{H}}_B}{{\mb{q}}_B}} \right\|}^2}}}{{{{\left\| {{{\mb{H}}_A}{{\mb{q}}_A}} \right\|}^2} + {{\left\| {{{\mb{H}}_B}{{\mb{q}}_B}} \right\|}^2}}},  \\
   & \quad \quad \quad \quad \text{if } N_A + N_B > N_R, \\
   & \frac{1}{2}\log_2 \frac{1}{{1 - {\left(\sigma_{\max}(\mb{U}_A^H \mb{U}_B)\right)^2}}},  \text{if }  N_A + N_B \leq N_R, \\
    \end{split}
    \right.,
 \end{eqnarray}
 where set $\mathcal{S}$ is $\{(\mb{q}_A, \mb{q}_B): \exists \beta \in \mathbb{R}, \beta {\mb{H}_A}{\mb{q}_A} = {\mb{H}_B}{\mb{q}_B} \\ \text{ and } \|\mb{q}_A\|^2 \leq P_A, \|\mb{q}_B\|^2 \leq P_B\}$, $\sigma_{\max}(\mb{U}_A^H \mb{U}_B)$ is the maximum singular value of matrix $\mb{U}_A^H \mb{U}_B$, $\mb{U}_A \in \mathbb{C}^{N_R \times \min\{N_A, N_R\}}$ and $\mb{U}_B \in \mathbb{C}^{N_R \times \min\{N_B, N_R\}}$ are obtained from the QR decomposition of $\mb{H}_A$ and $\mb{H}_B$, respectively, i.e.,
 \begin{eqnarray}\label{eq_H_qr_decomposition}
   \mb{H}_i = \mb{U}_i \mb{R}_{i}, \quad i\in \{A,B\},
 \end{eqnarray}
 where $\mb{R}_{i} \in \mathbb{C}^{\min\{N_R, N_i\} \times N_i }$ are upper triangle matrices.
\end{proposition}

\begin{IEEEproof}
We first prove the following fact:

When $P_R \to \infty$, the information rate from $i$ to $\bar{i}$ in two-phase two-way relay scheme is
\begin{eqnarray}\label{eq_Rate_2P_P_R_infty}
      \lim \limits_{P_R \to \infty} R_{i\bar{i}}^{2P} =  \frac{1}{2} \log_2 \left( {1 + \mb{q}_i^H \mb{H}_i^H {\mb{H}_i} {\mb{q}_i}} \right),
\end{eqnarray}

  To prove \eqref{eq_Rate_2P_P_R_infty}, we first plug in the optimal structure of $\mb{F}$ to \eqref{eq_2P_R_AB} and let $\mb{F} = t \mb{V} \mb{A} \mb{U}^H$ where $t$ is a real number. When $P_R \to \infty$, we just let $t \to \infty$. Thus,
\small
\begin{eqnarray*}
    & &\mb{q}_i^H \mb{H}_i^H{\mb{F}^H} \mb{G}_{\bar{i}}^H \mb{K}_{\bar{i}}^{-1} {\mb{G}_{\bar{i}}} \mb{F}{\mb{H}_i} {\mb{q}_i} \\
    &=& {\mb{q}}_i^H{\mb{H}}_i^H t {\mb{U}} {{\mb{A}}^H} {{\mb{V}}^H}{\mb{G}}_{\bar{i}}^H \left( t^2 \mb{G}_{\bar{i}} \mb{V} \mb{A} \mb{A}^H \mb{V}^H \mb{G}_{\bar{i}}^H + \mb{I}\right)^{-1} \cdot \\
    & &  {{\mb{G}}_{\bar{i}}} t \mb{V} \mb{A} {{\mb{U}}^H}{{\mb{H}}_i}{{\mb{q}}_i} \\
    &\stackrel{(a)}=& \mb{q}_i^H \mb{H}_i^H \mb{U} \left(\mb{I} - \left( \mb{I} + t^2 \mb{A}^H \mb{V}^H \mb{G}_{\bar{i}}^H \mb{G}_{\bar{i}} \mb{V} \mb{A}  \right)^{-1} \right) \mb{U}^H \mb{H}_i \mb{q}_i\\
    &=&\mb{q}_i^H \mb{H}_i^H \mb{U} \mb{U}^H \mb{H}_i \mb{q}_i -
    \mb{q}_i^H \mb{H}_i^H \mb{U} \left(\mb{I} + t^2 \mb{A}^H \mb{V}^H \mb{G}_{\bar{i}}^H \mb{G}_{\bar{i}} \mb{V} \mb{A}  \right)^{-1} \cdot\\
    & &  \mb{U}^H \mb{H}_i \mb{q}_i \\
    &\stackrel{(b)}=&  \mb{q}_i^H \mb{H}_i^H \mb{H}_i \mb{q}_i -
    \mb{q}_i^H \mb{H}_i^H \mb{U} \left(\mb{I} + t^2 \mb{A}^H \mb{V}^H \mb{G}_{\bar{i}}^H \mb{G}_{\bar{i}} \mb{V} \mb{A}  \right)^{-1} \mb{U}^H \mb{H}_i \mb{q}_i
\end{eqnarray*}
\normalsize
where $(a)$ is from the matrix inverse lemma and $(b)$ is from QR decomposition \eqref{QR_2_phase_Hq}. Since nodes $A$, $B$ and $R$ all have multiple antennas, we have $\mathrm{Rank}\left(\mb{G}_{\bar{i}}\right) \geq 2$ with probability 1 as every element of $\mb{G}_{\bar{i}}$ are drawn from continuous distribution. Therefore, it is always possible to find $\mb{A}$ such that $\left(\mb{A}^H \mb{V}^H \mb{G}_{\bar{i}}^H \mb{G}_{\bar{i}} \mb{V} \mb{A}\right) \in \mathbb{C}^{2\times 2}$ is positive definite matrix. Hence, the eigenvalue of $\left(\mb{I} + t^2 \mb{A}^H \mb{V}^H \mb{G}_{\bar{i}}^H \mb{G}_{\bar{i}} \mb{V} \mb{A}  \right)$ approaches positive infinity when $t \to \infty$.
As a result, the term $\mb{q}_i^H \mb{H}_i^H \mb{U} \left(\mb{I} + t^2 \mb{A}^H \mb{V}^H \mb{G}_{\bar{i}}^H \mb{G}_{\bar{i}} \mb{V} \mb{A}  \right)^{-1} \mb{U}^H \mb{H}_i \mb{q}_i$ approaches zero and we obtain that when ${P_R \to \infty}$, $R_{i\bar{i}}^{2P} \geq  \frac{1}{2} \log_2 \left( {1 + \mb{q}_i^H \mb{H}_i^H {\mb{H}_i} {\mb{q}_i}} \right)$. In addition, it is easy to see that $\lim \limits_{P_R \to \infty} R_{i\bar{i}}^{2P} \leq  \frac{1}{2} \log_2 \left( {1 + \mb{q}_i^H \mb{H}_i^H {\mb{H}_i} {\mb{q}_i}} \right)$. Therefore, we obtain \eqref{eq_Rate_2P_P_R_infty}.

Substituting \eqref{eq_2P_AB_R} and \eqref{eq_Rate_2P_P_R_infty}  into \eqref{eq_2P_R_s}, we obtain the achievable sum-rate as
\begin{eqnarray}\label{eq_2P_secrey_rate_P_R_infty}
   \lim \limits_{P_R \to \infty} \mathcal{R}_s^{2P} &=& \frac{1}{2}\log_2 \frac{1}{{1 - f(\mb{q}_A, \mb{q}_B)}}
\end{eqnarray}
where
\begin{eqnarray}\label{eq_f}
  f(\mb{q}_A, \mb{q}_B) &\triangleq & \frac{{{{| {{\mb{q}}_A^H{\mb{H}}_A^H{{\mb{H}}_B}{{\mb{q}}_B}} |}^2}}}{{\left( {1 + {{\left\| {{{\mb{H}}_B}{{\mb{q}}_B}} \right\|}^2}} \right)\left( {1 + {{\left\| {{{\mb{H}}_A}{{\mb{q}}_A}} \right\|}^2}} \right)}}.
\end{eqnarray}

From \eqref{eq_2P_secrey_rate_P_R_infty}, we see that to maximize
$\lim_{P_R \to \infty} \mathcal{R}_s^{2P}$, we should maximize $f(\mb{q}_A, \mb{q}_B)$.
An upper bound of $f(\mb{q}_A, \mb{q}_B)$ is,
\begin{eqnarray} \label{eq_f_bar}
  f(\mb{q}_A, \mb{q}_B)< \frac{{{{| {{\mb{q}}_A^H{\mb{H}}_A^H{{\mb{H}}_B}{{\mb{q}}_B}} |}^2}}}{{{{\left\| {{{\mb{H}}_B}{{\mb{q}}_B}} \right\|}^2}{{\left\| {{{\mb{H}}_A}{{\mb{q}}_A}} \right\|}^2}}} \leq 1,
\end{eqnarray}
and this upper bound can be approached when $P_A \to \infty$ and $P_B \to \infty$, i.e.,
\begin{eqnarray}
   \bar{f}(\mb{q}_A, \mb{q}_B) \triangleq \lim_{P_A \to \infty, P_B \to \infty}f(\mb{q}_A, \mb{q}_B)= \frac{{{{| {{\mb{q}}_A^H{\mb{H}}_A^H{{\mb{H}}_B}{{\mb{q}}_B}} |}^2}}}{{{{\left\| {{{\mb{H}}_B}{{\mb{q}}_B}} \right\|}^2}{{\left\| {{{\mb{H}}_A}{{\mb{q}}_A}} \right\|}^2}}}.
\end{eqnarray}
Therefore, the problem is transformed to maximize $\bar{f}(\mb{q}_A, \mb{q}_B)$, which is to find two vectors with the minimum angle from the column spaces of $\mb{U}_A$ and $\mb{U}_B$.

For the case $N_A + N_B > N_R$, with probability one, we can find $\mb{q}_A$ and $\mb{q}_B$ such that
\begin{equation}\label{eq_align}
  \beta \mb{H}_A \mb{q}_A= \mb{H}_B \mb{q}_B
\end{equation}
where $\beta$ can be an arbitrary non-zero real number. Under this condition, $\bar{f}(\mb{q}_A, \mb{q}_B)$ can take its maximum value of $1$ in \eqref{eq_f_bar}\footnote{An algorithm to find $\mb{q}_A$ and $\mb{q}_B$ was shown in \cite[Lemma 1]{Lee_Namyoon_IT10}.}.
Therefore, substituting the condition \eqref{eq_align} into \eqref{eq_2P_secrey_rate_P_R_infty}, we obtain
\begin{eqnarray*}
    \lim \limits_{P_R \to \infty} \mathcal{R}_s^{2P} &=& \frac{1}{2}\log_2 \frac{1}{{1 - f(\mb{q}_A, \mb{q}_B)}} \\
    &=& \frac{1}{2} \log_2 \frac{\left(1+\left\|\mb{H}_A \mb{q}_A\right\|^2\right)\left(1+\left\|\mb{H}_B \mb{q}_B\right\|^2\right)}{1+{\left\|\mb{H}_A \mb{q}_A\right\|}^2 + \left|\mb{H}_B \mb{q}_B\right|^2}\\
    &\approx& \frac{1}{2}{\log _2}\frac{{{{\left\| {{{\mb{H}}_A}{{\mb{q}}_A}} \right\|}^2}{{\left\| {{{\mb{H}}_B}{{\mb{q}}_B}} \right\|}^2}}}{{{{\left\| {{{\mb{H}}_A}{{\mb{q}}_A}} \right\|}^2} + {{\left\| {{{\mb{H}}_B}{{\mb{q}}_B}} \right\|}^2}}} \\
    & & \text{if } P_A \rightarrow \infty, P_B \rightarrow \infty.
 \end{eqnarray*}
At last, we maximize over the possible alignment directions and obtain the first part of Proposition \ref{proposition_2P_P_R_infty_P_i_infty}.

On the other hand, if $N_A + N_B \leq N_R$, we have,
\begin{eqnarray}
& &\left\| {{\mb{q}}_B^H{\mb{H}}_B^H{{\mb{H}}_A}{{\mb{q}}_A}} \right\| \nonumber\\
&\stackrel{(a)}=& \left\| {{\mb{q}}_B^H{\mb{R}}_B^H{\mb{U}}_B^H{{\mb{U}}_A}{{\mb{R}}_A}{{\mb{q}}_A}} \right\| \nonumber \\
&\stackrel{(b)}\le & {\sigma _{\max }}\left( {{\mb{U}}_B^H{{\mb{U}}_A}} \right)\left\| {{{\mb{R}}_A}{{\mb{q}}_A}} \right\|\left\| {{{\mb{R}}_B}{{\mb{q}}_B}} \right\| \nonumber\\
&\stackrel{(c)}=& {\sigma _{\max }}\left( {{\mb{U}}_B^H{{\mb{U}}_A}} \right)\left\| {{{\mb{U}}_A}{{\mb{R}}_A}{{\mb{q}}_A}} \right\|\left\| {{{\mb{U}}_B}{{\mb{R}}_B}{{\mb{q}}_B}} \right\| \nonumber \\
&\stackrel{(d)}=& {\sigma _{\max }}\left( {{\mb{U}}_B^H{{\mb{U}}_A}} \right)\left\| {{{\mb{H}}_A}{{\mb{q}}_A}} \right\|\left\| {{{\mb{H}}_B}{{\mb{q}}_B}} \right\| \label{eq_P1_proof}
\end{eqnarray}
where $(a)$ and $(d)$ are from \eqref{eq_H_qr_decomposition}, $(b)$ is from the singular value decomposition of $\mb{U}_B^H \mb{U}_A$ and the equality can be achieved by letting ${{\mb{q}}_A} = {\mb{R}}_A^{ - 1}{{\mb{\psi }}_{\max }}\left( {{\mb{U}}_A^H{{\mb{U}}_B}{\mb{U}}_B^H{{\mb{U}}_A}} \right)$ and ${{\mb{q}}_B} = {\mb{R}}_B^{ - 1}{{\mb{\psi }}_{\max }}\left( {{\mb{U}}_B^H{{\mb{U}}_A}{\mb{U}}_A^H{{\mb{U}}_B}} \right)$ where the upper triangle matrices $\mb{R}_i \in \mathbb{C}^{N_i \times N_i}$ are invertible, and $(c)$ is from ${\mb{q}}_i^H{\mb{R}}_i^H{{\mb{R}}_i}{{\mb{q}}_i} = {\mb{q}}_i^H{\mb{R}}_i^H{\mb{U}}_i^H{{\mb{U}}_i}{{\mb{R}}_i}{{\mb{q}}_i}$. Substituting \eqref{eq_P1_proof} back to \eqref{eq_2P_secrey_rate_P_R_infty}, we obtain the second part of Proposition \ref{proposition_2P_P_R_infty_P_i_infty}.

{\color{black}Notice that we always have ${\sigma _{\max }}\left( {{\mb{U}}_B^H{{\mb{U}}_A}} \right) < 1$ when $N_A+N_B \leq N_R$. The proof is as follows.

First, as $\left\| {{\mb{q}}_B^H{\mb{H}}_B^H{{\mb{H}}_A}{{\mb{q}}_A}} \right\| \leq \left\| {{{\mb{H}}_A}{{\mb{q}}_A}} \right\|\left\| {{{\mb{H}}_B}{{\mb{q}}_B}} \right\| $ and the equality in $(b)$ can be achieved, we know that ${\sigma _{\max }}\left( {{\mb{U}}_B^H{{\mb{U}}_A}} \right) \leq 1$.
Second, if ${\sigma _{\max }}\left( {{\mb{U}}_B^H{{\mb{U}}_A}} \right) = 1 $, there is an intersection subspace between the space  ${\rm span}(\mb{H}_A)$ and ${\rm span}(\mb{H}_B)$ such that $\beta {{{\mb{H}}_A}{{\mb{q}}_A}}= {{{\mb{H}}_A}{{\mb{q}}_A}}$ where $\beta$ is a real number. However, according to \textit{dimension theorem} \cite{Strang_04} and because the entries of the channel matrices are generated from continuous distribution, we have
\begin{eqnarray*}
  & & {\rm dim}({\rm span}(\mb{H}_A) \cap {\rm span}(\mb{H}_A)) \\
  &=& {\rm dim}({\rm span}(\mb{H}_A)) + {\rm dim}({\rm span}(\mb{H}_B)) - {\rm dim}({\rm span}([\mb{H}_A, \mb{H}_B])) \\
  &=& N_A + N_B - (N_A + N_B) \\
  &=& 0.
\end{eqnarray*}
Consequently, there is no intersection subspace and we have $\sigma_{\max}\left(\mb{U}_B^H \mb{H}_A\right)<1$.}

Thus, the proof of Proposition \ref{proposition_2P_P_R_infty_P_i_infty} is completed.
\end{IEEEproof}

\begin{figure}[t]
\begin{centering}
\includegraphics[scale=.30]{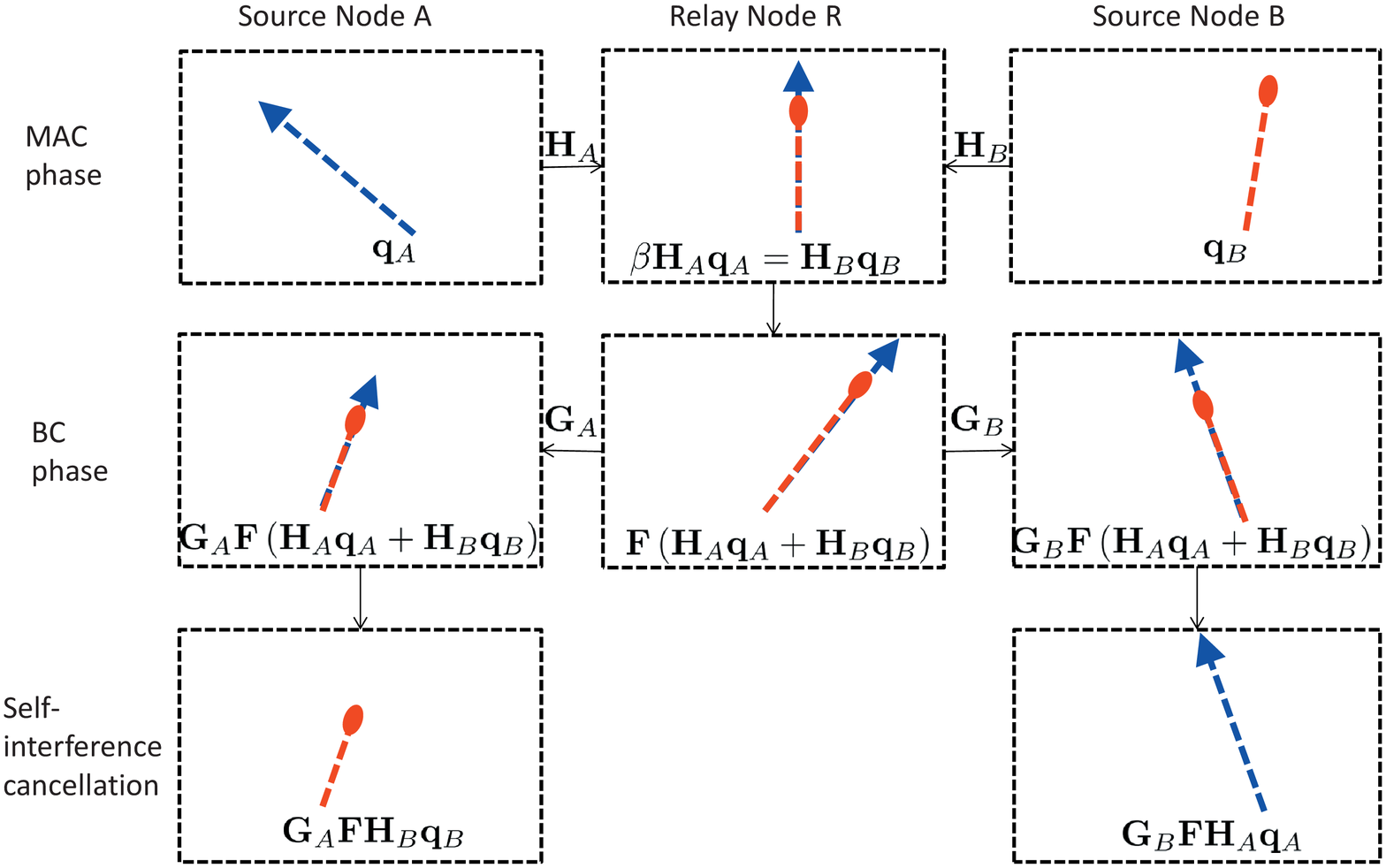}
\vspace{-0.1cm}
\centering
 \caption{The signal vectors of the two-phase two-way relaying scheme.}\label{fig:Signal_alignment}
\end{centering}
\vspace{-0.3cm}
\end{figure}

Proposition \ref{proposition_2P_P_R_infty_P_i_infty} is essentially similar as the so-called signal alignment. In \cite{Lee_Namyoon_IT10}, this technique was first proposed to achieve the degrees of freedom of the MIMO Y channel which is a generalized two-way relay channel with three users. The key idea of the signal alignment is to align the two desired signal vectors coming from two users at the receiver of the relay to jointly perform detection and encoding for network coding. Specifically, if $N_A+N_B>N_R$, there is intersection subspace between the column spaces of $\mb{H}_i$ with probability one and thus there exists $\beta \in \mathbb{R}$ such that \eqref{eq_align} holds. As illustrated in Fig. \ref{fig:Signal_alignment}, the secure beamformers at the two source nodes are chosen such that the two received signals align in the same direction at the relay node. Intuitively, aligning the signal vectors at the relay node will hinder the relay node decode the source messages and make the system more secure.  After self-interference cancellation, the two source nodes will obtain the desired signal. The maximum secrecy sum rate goes to infinity as the source powers approach infinity.
On the other hand, if $N_A+N_B\leq N_R$, there is no intersection subspace with probability one and there is an upper bound for the maximum secrecy sum rate. Specifically, $\mb{U}_i$ is the orthonormal basis of the column space of $\mb{H}_i$. {\color{black}Thus, $\arccos \left( \sigma_{\max}\left({\mb{U}_A^H \mb{U}_B} \right)\right)$, is the minimum angle between any two vectors from the respect two column spaces. Actually, it is called the minimum principal angle of these two subspaces \cite{Golub_12_matrix}.}

\begin{proposition}[3P]
\label{proposition_3P_P_R_infty_P_i_infty}
    When $P_R \to \infty$, $P_A \to \infty$ and $P_B \to \infty$,
  the maximum secrecy sum rate of the three-phase two-way relay scheme is,
  \begin{eqnarray}
  \mathcal{R}_{\max}^{3P} \approx
 \sum\limits_{i \in \left\{ {A,B} \right\}} {{\Theta _i}},
 \end{eqnarray}
 where
 \begin{eqnarray*}
 & \Theta _i \in \Bigg[ \frac{1}{3}{{\left[ {{{\log }_2}\left( {\frac{1}{2} + {\lambda _{\max }}\left( {{\mb{T}}_i^H{{\mb{T}}_i},{\mb{H}}_i^H{{\mb{H}}_i}} \right)} \right)} \right]}^ + }, \\
 & \quad \quad \frac{1}{3}{{\left[ {{{\log }_2}\left( {1 + {\lambda _{\max }}\left( {{\mb{T}}_i^H{{\mb{T}}_i},{\mb{H}}_i^H{{\mb{H}}_i}} \right)} \right)} \right]}^ + } \Bigg],
 \end{eqnarray*}
 if $N_i \leq N_R$ and
 \begin{equation*}
 {\Theta _i} = \frac{1}{3}\left[ {{{\log }_2}\left( {\frac{3}{2}{P_i}} \right) + {{\log }_2}\left( {{\lambda _{\max }}\left( {{\mb{H}}_i^\mathcal{N}{\mb{T}}_i^H{{\mb{T}}_i}{\mb{H}}_i^\mathcal{N}} \right)} \right)} \right],
\end{equation*}
if ${N_i} > {N_R}$.
\end{proposition}
\begin{IEEEproof}
    See Appendix \ref{app_3P_P_R_infty_P_i_infty}.
\end{IEEEproof}
{\color{black}Proposition \ref{proposition_3P_P_R_infty_P_i_infty} shows that the secrecy sum-rate of the three-phase scheme will reach a floor if the untrusted relay has more antennas than the two source nodes.}

\begin{proposition}[DT]
\label{proposition_DT_P_R_infty_P_i_infty}
    When $P_A \to \infty$ and $P_B \to \infty$,
  the maximum secrecy sum rate of the two-way direct transmission scheme is
  \begin{eqnarray}
    \mathcal{R}_{\max}^{DT} \approx
 \sum\limits_{i \in \left\{ {A,B} \right\}} {{\Omega _i}} ,
 \end{eqnarray}
 where
 \begin{eqnarray*}
  {\Omega _i} = \left\{ \begin{split}
  & \frac{1}{2}{\left[ {\log_2 {\lambda _{\max }}\left( {{\mb{T}}_i^H{{\mb{T}}_i},{\mb{H}}_i^H{{\mb{H}}_i}} \right)} \right]^ + },\quad & &\text{if } {N_i} \leq {N_R}  \\
  & \frac{1}{2}{\left[ {\log_2 {P_i} + \log_2 {\lambda _{\max }}\left( {{\mb{H}}_i^{\mathcal{N}}{\mb{T}}_i^H{{\mb{T}}_i}{\mb{H}}_i^{\mathcal{N}}} \right)} \right] }, & & \text{if } {N_i} > {N_R}  \\
\end{split}  \right. ,
  \end{eqnarray*}
  and the optimal beamforming $\mb{q}_i^{DT}$ is given in \eqref{dt_Beamforming}.
\end{proposition}
\begin{IEEEproof}
    This lemma is based on \cite[Lemma 7]{Jeong_TSP12}. Here, we assume that the entries of the channel matrices are generated from continuous distribution. As a result, $\mathrm{Rank}(\mb{H}^{m\times n}) \geq \min\{m,n\}$ with probability one. In addition, the condition $\mb{H}_i^{\mathcal{N}} \mb{T}_i^H \neq \mb{0}$ in \cite[Lemma 7]{Jeong_TSP12} is also satisfied with probability one.
\end{IEEEproof}

{\color{black}As shown in Proposition \ref{proposition_DT_P_R_infty_P_i_infty}, the secrecy sum-rate of the direct transmission scheme will also reach a floor if untrusted relay has more antennas than the source nodes. This is similar to the three-phase case.}

From Proposition \ref{proposition_2P_P_R_infty_P_i_infty}, \ref{proposition_3P_P_R_infty_P_i_infty}, and \ref{proposition_DT_P_R_infty_P_i_infty}, we find that the asymptotic comparison among these three schemes depend not only on the antenna numbers $N_A$, $N_B$, $N_R$ but also on specific channel realizations. In the following, we only present the comparison results in two cases.

\begin{corollary}
\label{corollary_P_R_infty_P_i_infty_case_1}
     When $P_R\to \infty$, and  $N_A \leq N_R$, $N_B \leq N_R$, $N_A + N_B > N_R$, the maximum of secrecy sum rate of the two-phase two-way relay scheme keeps increasing when the two source powers $P_A$ and $P_B$ increase while the maximum of secrecy sum rates of two-way direct transmission and three-phase scheme both approach constants. Thus, we have
    \begin{eqnarray}\label{eq_P_R_infty_P_i_infty_case_2}
        \mathcal{R}_{\max}^{2P} \geq
      \max \left\{ \mathcal{R}_{\max}^{DT}, \mathcal{R}_{\max}^{3P} \right\}.
    \end{eqnarray}
\end{corollary}

\begin{IEEEproof}
  It can be easily verified from Proposition \ref{proposition_2P_P_R_infty_P_i_infty}, \ref{proposition_3P_P_R_infty_P_i_infty} and \ref{proposition_DT_P_R_infty_P_i_infty}.
\end{IEEEproof}

\begin{remark}
As shown in \cite{Jeong_TSP12,Khisti_IT10a} for two-way direct transmission, in the infinite power case, the infinite maximum secrecy sum rate needs $N_A>N_R$ or $N_B>N_R$. Proposition \ref{proposition_2P_P_R_infty_P_i_infty} reveals that with the signal alignment techniques at the untrusted relay, the infinite maximum secrecy sum rate only needs $N_A+N_B>N_R$, which lowers the requirement of the numbers of antennas at the two sources. The result clearly demonstrates the benefits of signal alignment for physical layer security, which is the unique feature in two-way relaying.
\end{remark}

\begin{corollary}
\label{corollary_P_R_infty_P_i_infty_case_2}
    When $P_R\to \infty$, $P_A\to \infty$, $P_B\to \infty$ and $N_A > N_R$, $N_B > N_R$,
    \begin{equation}\label{eq_P_R_infty_P_i_infty_case_2}
       \mathcal{R}_{\max}^{DT} \geq \mathcal{R}_{\max}^{3P}.
    \end{equation}
\end{corollary}
\begin{IEEEproof}
    When $N_i > N_R$, we have
    \begin{eqnarray}
 \mathcal{R}_{\max }^{DT} = \sum\limits_{i \in \left\{ {A,B} \right\}} {\frac{1}{2}\left[ {{{\log }_2}{P_i} + \mathcal{O}\left( {{{\log }_2}{P_i}} \right)} \right]} \label{eq_DT_dof}\\
 \mathcal{R}_{\max }^{3P} = \sum\limits_{i \in \left\{ {A,B} \right\}} {\frac{1}{3}\left[ {{{\log }_2}{P_i} + \mathcal{O}\left( {{{\log }_2}{P_i}} \right)} \right]}\label{eq_3P_dof}
    \end{eqnarray}
    where the order notation $\mathcal{O}\left(P\right)$ means that $\mathcal{O}\left( P \right) \left.\right/ P  \to 0$ as $P \to \infty$. Thus, the Corollary \ref{corollary_P_R_infty_P_i_infty_case_2} follows.
\end{IEEEproof}

{\color{black}From this Corollary we see that when the number of antennas at each source node is larger than the number of antennas at the relay node, direct  transmission performs better than the three-phase two-way relaying at high SNR.}

\subsection{The Case of High Relay Power and Low Source Powers}
\label{subsection_P_i_0}

\begin{proposition}[2P]
\label{proposition_2P_P_R_infty_P_i_0}
  When $P_R \to \infty$, $P_A \to 0$ and $P_B \to 0$, the optimal source beamforming vectors of the two-phase two-way relay scheme are
  \begin{eqnarray}
    \mb{q}_A &=& \frac{\sqrt{P_A} \pmb{\psi}_{\max} \left(
\mb{H}_A^H  \mb{H}_B \mb{H}_B^H \mb{H}_A \right)}{\|\pmb{\psi}_{\max} \left(
\mb{H}_A^H  \mb{H}_B \mb{H}_B^H \mb{H}_A \right) \|}, \\
    \mb{q}_B &=& \frac{\sqrt{P_B} \pmb{\psi}_{\max} \left(
\mb{H}_B^H  \mb{H}_A \mb{H}_A^H \mb{H}_B \right)}{\|\pmb{\psi}_{\max} \left(
\mb{H}_B^H  \mb{H}_A \mb{H}_A^H \mb{H}_B \right) \|},
  \end{eqnarray}
  and the maximum secrecy sum rate is
  \begin{eqnarray}
    \mathcal{R}_{\max}^{2P} \approx \frac{1}{2 \ln 2}{P_A}{P_B}{\lambda _{\max}}\left( {\mb{H}_A^H \mb{H}_B \mb{H}_B^H \mb{H}_A } \right).
  \end{eqnarray}

\end{proposition}
\begin{IEEEproof}
    See Appendix \ref{app_2P_P_R_infty_P_i_0}.
\end{IEEEproof}
{\color{black}Note that $\mb{q}_A$ and $\mb{q}_B$ are determined by the concatenated channel $\mb{H}_A^{H} \mb{H}_B$. }

\begin{proposition}[3P]
\label{proposition_3P_P_R_infty_P_i_0}
    When $P_R \to \infty$, $P_A \to 0$ and $P_B \to 0$,
  the maximum secrecy sum rate of the three-phase two-way relay scheme satisfies
  \begin{eqnarray*}
    \frac{1}{2\ln 2}\sum\limits_{i \in \left\{ {A,B} \right\}} {{{\left[ {{P_i}{\lambda _{\max }}\left( {{\mb{T}}_i^H{{\mb{T}}_i} - \frac{1}{2}{\mb{H}}_i^H{{\mb{H}}_i}} \right)} \right]}^ + }} \\
    \leq \mathcal{R}_{\max}^{3P} \le \frac{1}{2\ln 2}\sum\limits_{i \in \left\{ {A,B} \right\}} {{P_i}{\lambda _{\max }}\left( {{\mb{T}}_i^H{{\mb{T}}_i}} \right)} .
  \end{eqnarray*}
\end{proposition}
\begin{IEEEproof}
    Substituting the above upper bound and lower bound of $\lim_{P_R \to \infty} \mathcal{R}_{i\bar{i}}^{3C}$ given in \eqref{eq_3P_UB_LB_rate} into the \eqref{eq_3P_secrecy_sum_rate}, we can easily prove Proposition \ref{proposition_3P_P_R_infty_P_i_0}.
\end{IEEEproof}

\begin{proposition}[DT]
\label{proposition_DT_P_R_infty_P_i_0}
    When $P_A \to 0$ and $P_B \to 0$,
  the maximum secrecy sum rate of the two-way direct transmission scheme is,
  \begin{eqnarray*}
     \mathcal{R}_{\max}^{DT} \approx
\frac{1}{2 \ln 2}\sum\limits_{i \in \left\{ {A,B} \right\}} {{{\left[ {{P_i}{\lambda _{\max }}\left( {{\mb{T}}_i^H{{\mb{T}}_i} - {\mb{H}}_i^H{{\mb{H}}_i}} \right)} \right]}^ + }}
  \end{eqnarray*}
  and the optimal beamforming $\mb{q}_i^{DT}$ are given in \eqref{dt_Beamforming}.
 \end{proposition}
\begin{IEEEproof}
    It is easily obtained from \eqref{nc_rate} or \cite[Lemma 6]{Jeong_TSP12}.
\end{IEEEproof}

{\color{black} We find that different from the two-phase scheme, the secrecy sum rates of the direct transmission and the three-phase scheme are closely related to the term ${\mb{T}_i^H}{\mb{T}_i}-\alpha {\mb{H}_i^H}{\mb{H}_i}$ ($\alpha=0,1,\frac{1}{2}$).}

\begin{corollary}
\label{corollary_P_R_infty_P_i_0}
    When $P_R \to \infty$, $P_A \to 0$ and $P_B \to 0$,
    we have
    \begin{eqnarray*}
      \mathcal{R}_{\max}^{3P} \geq \mathcal{R}_{\max}^{DT} \geq \mathcal{R}_{\max}^{2P}.
    \end{eqnarray*}
\end{corollary}
\begin{IEEEproof}
  This corollary can be easily obtained from Proposition \ref{proposition_2P_P_R_infty_P_i_0}, \ref{proposition_3P_P_R_infty_P_i_0} and \ref{proposition_DT_P_R_infty_P_i_0}. Since
   $\mb{H}_i^H \mb{H}_i$ are positive semidefinite matrices, $\lambda _{\max } \left({{\mb{T}}_i^H{{\mb{T}}_i} - \frac{1}{2}{\mb{H}}_i^H{{\mb{H}}_i}} \right) \geq \lambda _{\max } \left({{\mb{T}}_i^H{{\mb{T}}_i}-{\mb{H}}_i^H{{\mb{H}}_i}} \right)$. Therefore,
  %
  the three-phase two-way relay scheme is better than direct transmission scheme. In addition, $\mathcal{R}_{\max}^{2P}$ approaches zero faster than the other two schemes. Thus, the proof of Corollary \ref{corollary_P_R_infty_P_i_0} is completed.
\end{IEEEproof}

{\color{black}This Corollary clearly suggests that when the two source powers are extremely low, it is the best to apply the three-phase two-way relay scheme for secure transmission.}

\subsection{The Case of Low Relay Power}
In this subsection, we present the asymptotic secrecy sum rate when relay power approaches zero.

First, we briefly show when the relay power $P_R\to0$, the maximum secrecy sum rate of the two-phase two-way relay scheme $\mathcal{R}_{\max}^{2P}$ approaches zero. As the relay power approaches zero, the information rate through the relay link goes to zero, which means that $\mathcal{R}_{AB}^{2P}+\mathcal{R}_{BA}^{2P}$ approaches zero. On the other hand, the information rate leaked to untrusted relay $\mathcal{R}_{R}^{2P}$ is not related to the relay power and does not approach zero. Therefore, the secrecy sum rate is zero when $P_R \to 0$.

\begin{corollary}
\label{corollary_P_R_0}
  When the relay power $P_R\to 0$,
  \begin{equation}
    \mathcal{R}_{\max}^{DT} \geq \mathcal{R}_{\max}^{3P} \geq \mathcal{R}_{\max}^{2P}.
  \end{equation}
\end{corollary}
\begin{IEEEproof}
    See Appendix \ref{app_proposition_P_R_0}.
\end{IEEEproof}

{\color{black}Corollary \ref{corollary_P_R_0} shows that the direct transmission is the best when the relay power is low. In the relay system without secrecy constraint, the similar conclusion hold \cite{Chen_Min_TWC10}.}

 we can now summarize the main comparison results in Table \ref{table_comparison}. Utilizing Table \ref{table_comparison}, we can choose the best transmission scheme under different scenarios.

 {\color{black}Note that besides the three schemes we considered in this work, four-phase one-way relay scheme  is also possible for secure bi-directional transmission.
 In this four-phase scheme, the conventional one-way relaying is used twice for communications as $A \rightarrow R \rightarrow B$ and $B \rightarrow R \rightarrow A$. It can be shown that this four-phase scheme is the best when $P_R \rightarrow \infty$, $P_A \rightarrow 0$ and $P_B \rightarrow 0$. For the other cases, this scheme is either suboptimal or the comparison depends on the channel realization.}

\begin{table*}
\centering\caption{\label{table_comparison} The comparison of the three schemes in terms of the maximum secrecy sum rate. (In the table, we use `DT', `2P', `3P' to denote the three schemes. And, A $>$ B means that scheme A is better than scheme B.)}
\begin{tabular}{|c|c|c|c|}
  \hline
  \multicolumn{3}{|c|}{Conditions} & Comparison  \\
  \hline
  \multirow{4}{*}{$P_R \to \infty$} & \multicolumn{2}{ c|}{$P_A \to 0$, $P_B \to 0$} & 3P $>$ DT $>$ 2P (Corollary 3)\\
  \cline{2-4}
  & \multirow{3}{*}{$P_A\to \infty$, $P_B\to \infty$} & $N_A + N_B > N_R$, $N_A \leq N_R$, $N_B \leq N_R$ & 2P $>$ DT and 2P $>$ 3P (Corollary 1)  \\
  \cline{3-4}
  & & $N_A > N_R$, $N_B>N_R$ & DT $>$ 3P (Corollary 2)\\
  \cline{3-4}
  & &Other cases &Channel dependent \\
  \hline
  \multicolumn{3}{|c|}{$P_R\to 0$} & DT $>$ 3P $>$ 2P (Corollary 4)\\
  \hline
\end{tabular}
\end{table*}

\section{Simulation Results and Discussions}
In this section, we perform simulation for all the cases discussed in section IV and V.
In the simulation, we assume that the channel reciprocity holds, i.e., $\mb{H}_A = \mb{G}_A^T$, $\mb{H}_B = \mb{G}_B^T$ and $\mb{T}_A = \mb{T}_B^T$. The following example of channel coefficients realization (every entry of the matrices is generated from $\mathcal{CN}(0,1)$ distribution) is used to show the asymptotical performance.
\small
\begin{eqnarray*}
\mb{\bar{H}}_A=   \left[
     \begin{matrix}
   0.2686 - 0.0965i   &0.1305 - 1.2373i   &0.6027 + 0.8313i \\
   0.9510 + 0.8678i  &-0.4450 + 0.2224i  &-0.4630 + 0.3531i \\
   0.4050 - 0.7642i  &-0.6673 - 0.7447i  &-0.0039 + 1.0646i  \\
  -0.9971 + 0.2578i  &-1.5888 - 0.9503i  &-0.4514 - 0.2944i  \\
  -1.1448 + 0.1069i  &-0.5209 - 0.0569i   &0.1598 + 0.0048i
     \end{matrix}
   \right]
\end{eqnarray*}

   \begin{eqnarray*}
\mb{\bar{H}}_B=   \left[
     \begin{matrix}
   0.3612 + 0.7099i  & -0.0464 - 1.1249i  & 0.6175 - 1.6643i \\
   0.6236 - 0.3490i  & 0.2193 + 0.8722i  &-0.8481 - 0.1791i \\
  -0.4814 - 0.3466i  & 0.2838 + 0.3014i  &-0.3683 + 1.6906i \\
  -0.2929 + 1.5306i  &-0.2643 + 0.8701i  &-1.6770 + 0.0192i \\
  -0.0722 + 0.1413i  & 0.1504 + 0.9271i  & 0.9011 - 0.3934i
     \end{matrix}
   \right]
   \end{eqnarray*}

   \begin{eqnarray*}
\mb{\bar{T}}_A=   \left[
     \begin{matrix}
    0.0538 + 1.3647i &  1.1100 - 0.5711i & -0.5226 - 0.0653i \\
   0.9241 - 0.9370i & -0.5684 - 1.1719i & -0.3993 - 0.6427i \\
  -0.0592 - 1.2997i & -0.9250 + 0.1194i  & 0.1469 + 0.4010i
     \end{matrix}
   \right]
   \end{eqnarray*}
   \normalsize
  If the channel matrix we need is smaller than the dimension of the above matrices, we simply choose the left upper part of the corresponding matrix. For instance, if $N_A=2$, $N_R=3$, we choose $\mb{H}_A = \mb{\bar{H}}_A(1:3,1:2)$.

\begin{figure}[t]
\begin{centering}
\includegraphics[scale=.40]{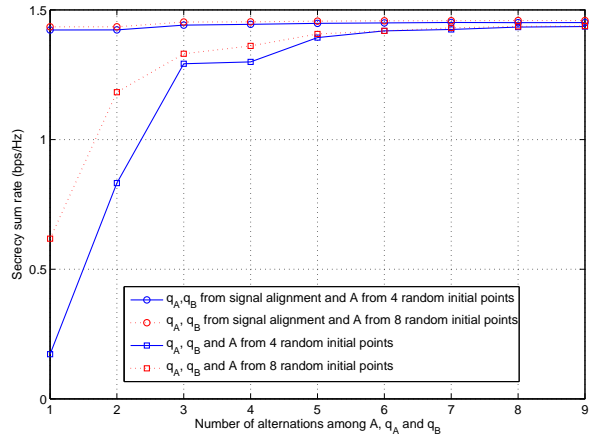}
\vspace{-0.1cm}
\centering
 \caption{ Convergence behaviour comparison of different initialization methods for Algorithm 1. $N_A=N_B=2$,  $N_R=3$, $P_R=30$ dB and $P_A=P_B=10$ dB.}\label{fig:Number_iteration}
\end{centering}
\vspace{-0.3cm}
\end{figure}

Note that Algorithm 1 and 2 are not guaranteed to find the optimal solution and the convergent point may be far from the optimal solution. A method to cope with this problem is to randomly generate multiple initializations and choose the one with the best performance. Fig. \ref{fig:Number_iteration} illustrates the convergence behavior of Algorithm 1 with different initializations. It is seen that when the initialization vectors $\mb{q}_A$ and $\mb{q}_B$ are chosen based on the signal alignment technique, the algorithm converges faster than the case of random generated vectors. Thus, in the rest of our simulation, we choose the asymptotic optimal beamforming vectors shown in Section IV as the initial points of $\mb{q}_A$ and $\mb{q}_B$.

\subsection{High Relay Power and High Source Powers}

\begin{figure}[t]
\begin{centering}
\includegraphics[scale=.40]{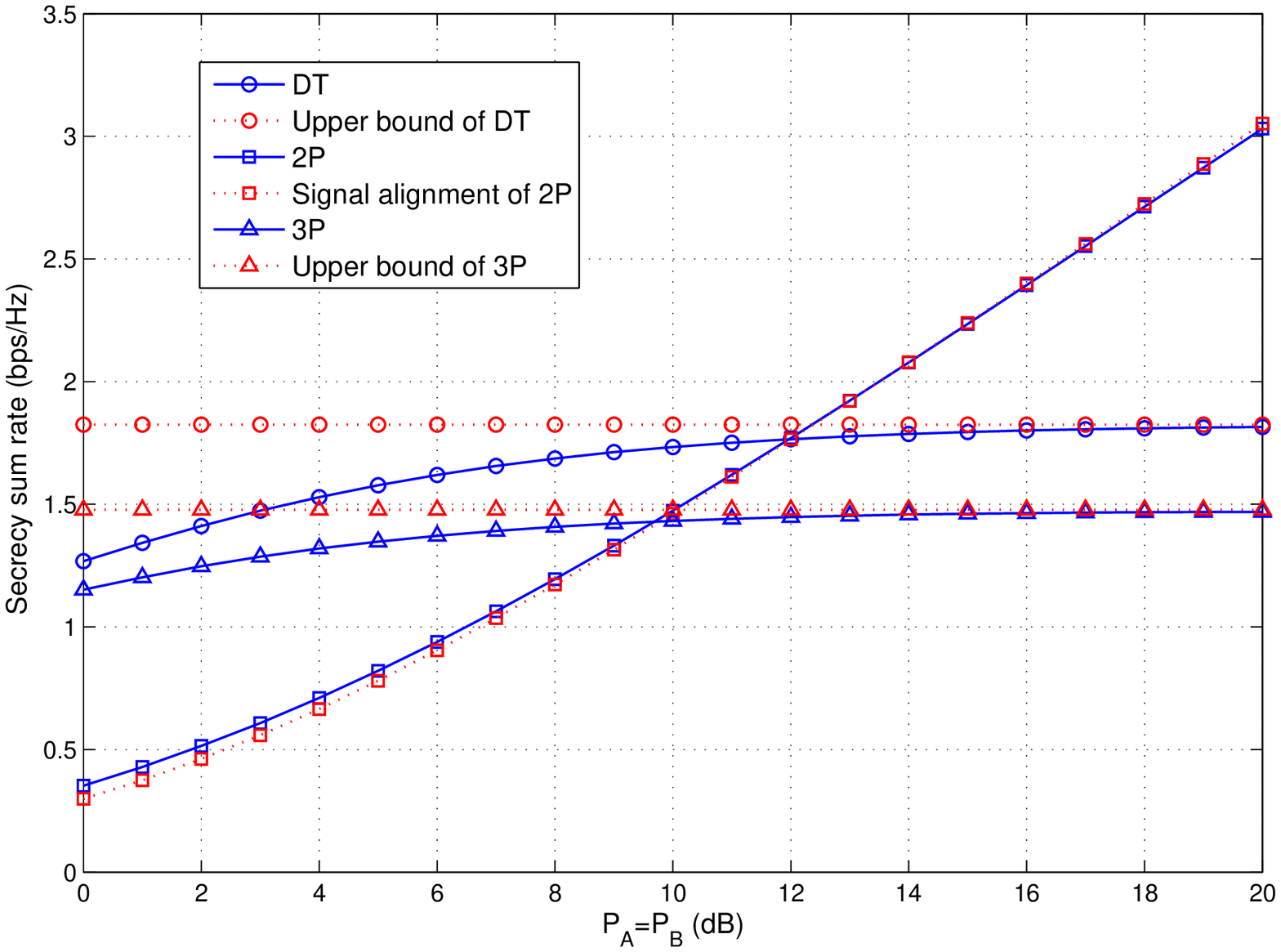}
\vspace{-0.1cm}
\centering
 \caption{Comparison of the three schemes in high power regimes when $N_A=2$, $N_R=3$, $N_B=2$ and $P_R =40$ dB.}\label{fig:Compare_2_3_2}
\end{centering}
\vspace{-0.3cm}
\end{figure}

\begin{figure}[t]
\begin{centering}
\includegraphics[scale=.40]{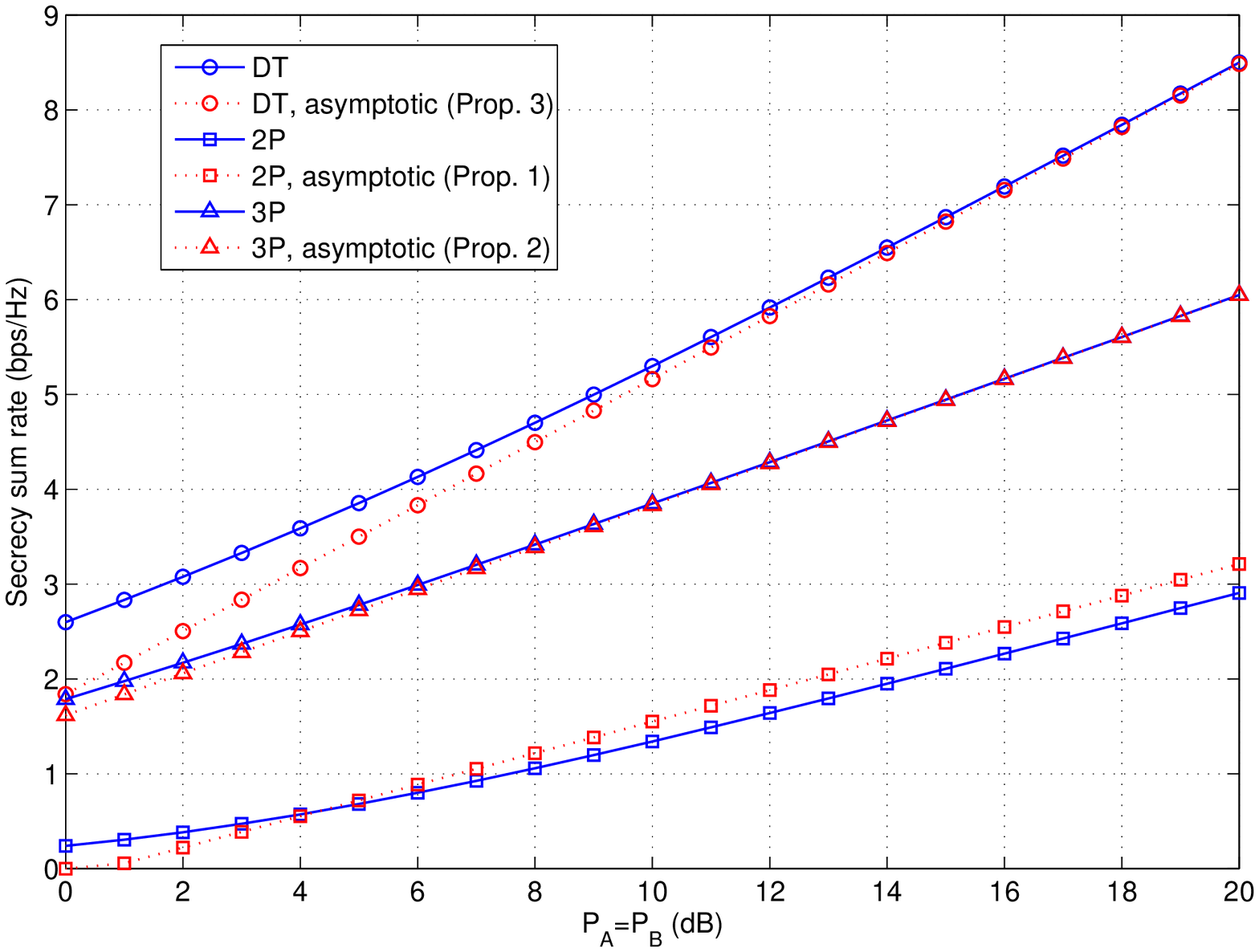}
\vspace{-0.1cm}
\centering
 \caption{Comparison of the three schemes in high power regimes when $N_A=3$, $N_R=2$, $N_B=3$ and $P_R =40$ dB.}\label{fig:Compare_3_2_3}
\end{centering}
\vspace{-0.3cm}
\end{figure}

\begin{figure}[t]
\begin{centering}
\includegraphics[scale=.40]{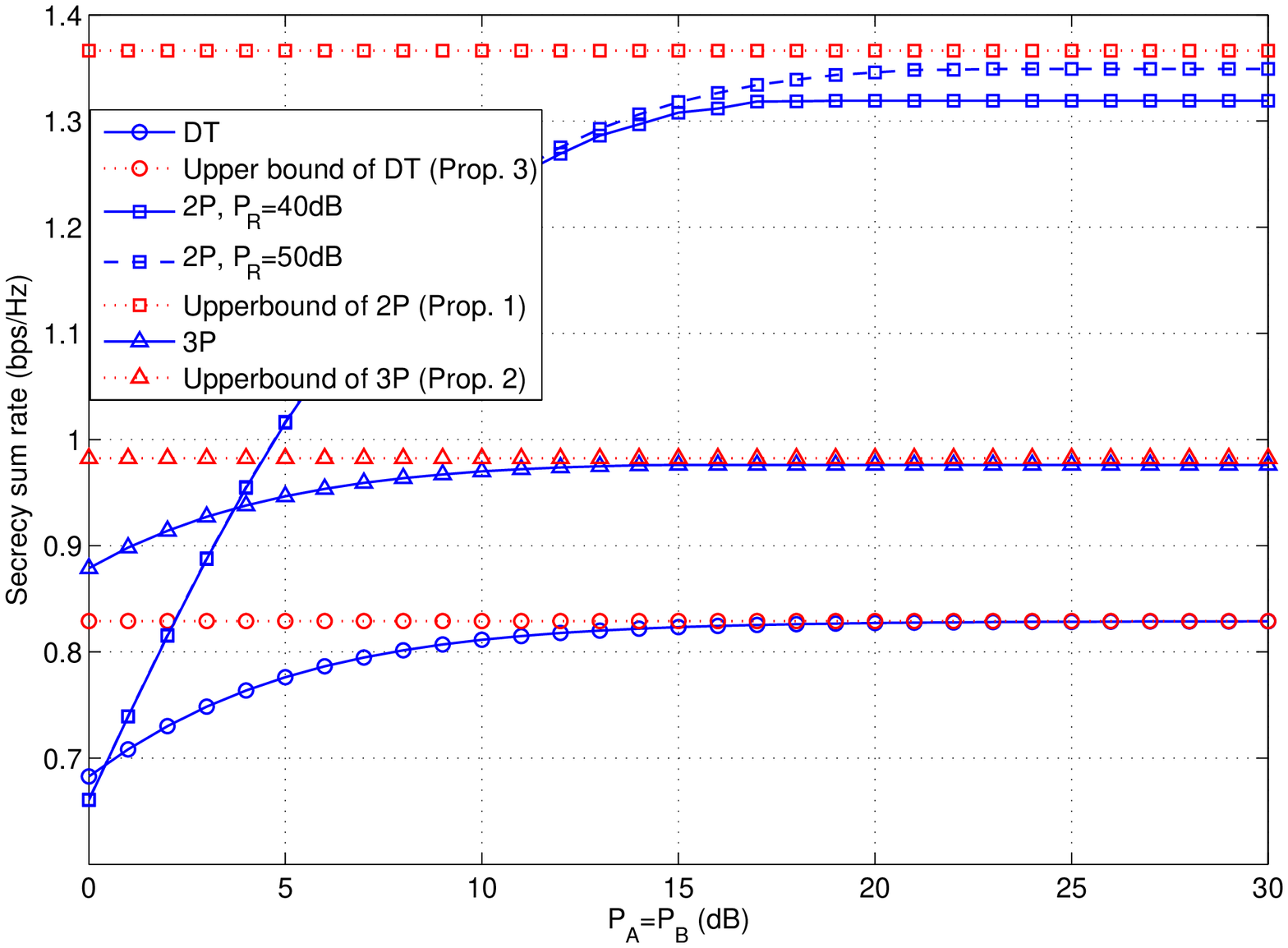}
\vspace{-0.1cm}
\centering
 \caption{Comparison of the three schemes in high power regimes when $N_A=2$, $N_R=5$, $N_B=2$ and $P_R =40$ dB.}\label{fig:Compare_2_5_2}
\end{centering}
\vspace{-0.3cm}
\end{figure}

{\color{black}Fig. \ref{fig:Compare_2_3_2}, \ref{fig:Compare_3_2_3} and \ref{fig:Compare_2_5_2} compare the secrecy sum rates obtained by different schemes. Here the relay power is fixed at $P_R = 40$dB, but the source powers are changing. The results for the two-phase and three-phase two-way relay scheme are obtained using the Algorithm 1 and 2 proposed in Section III. For the direct transmission, we use the closed-form expression \ref{nc_rate} given in Appendix \ref{app_DT}.}


Case $\left.1\right)$ $N_A=2$, $N_R=3$, $N_B=2$: {\color{black}This is an example of the case when $N_A<N_R$, $N_B<N_R$ and $N_A+N_B < N_R$. The curve for signal alignment of 2P is obtained by forcing $\beta \mb{H}_A \mb{q}_A=\mb{H}_B \mb{q}_B$.
Fig. \ref{fig:Compare_2_3_2} clearly verifies the importance of signal alignment for security  as analyzed in Corollary \ref{corollary_P_R_infty_P_i_infty_case_1}.} We see that in Fig.\ref{fig:Compare_2_3_2} the maximum secrecy sum rate of two-phase scheme goes to infinity with the increase of the source powers, while that of the other two schemes reach floors. Under this channel setup, the upper bound of the secrecy sum rate of the direct transmission scheme is about $1.82$bps/Hz  and that of three-phase scheme is $1.48$bps/Hz.

Case $\left.2\right)$ $N_A=3$, $N_R=2$, $N_B=3$: {\color{black}This is an example of the case when $N_A>N_R$ and $N_B>N_R$. As shown in Fig. \ref{fig:Compare_3_2_3}, the maximum secrecy sum rate for these schemes all approach to infinity as the powers increase. We find that the direct transmission scheme is the best. This agrees with our analysis in Corollary \ref{corollary_P_R_infty_P_i_infty_case_2}.} Actually, as shown in \eqref{eq_DT_dof} and \eqref{eq_3P_dof}, the degrees of freedom of the direct transmission scheme is one and the degrees of freedom of the three-phase scheme is $\frac{2}{3}$.  In this case, although the signal alignment of the two-phase scheme is feasible, the direct transmission scheme is better than the two-phase scheme.

Case $\left.3\right)$ $N_A=2$, $N_R=5$, $N_B=2$: This is the scenario when $N_A+N_B \leq N_R$. Under this condition, all the schemes have upper bounds for their secrecy sum rates. The comparison results are shown in Fig. \ref{fig:Compare_2_5_2}. It is shown that the two-phase scheme is the best. We also plot the curve for two-phase scheme when $P_R=50$dB. The curve can approach the upper bound more closely than the curve when $P_R=40$dB. This implies that to approach the upper bound given in \eqref{eq_2P_angle_bound}, we need the powers of all the three nodes go to infinity and the relay power should be much larger than the source powers. In this case, although the signal alignment of the two-phase scheme cannot be achieved, the two-phase scheme is better than the direct transmission scheme.

{\color{black}From Fig. \ref{fig:Compare_2_3_2} and Fig. \ref{fig:Compare_2_5_2}, we can see that increasing the number of antennas at the relay reduces the performance. This is in contrast  to the relay system without secrecy constraints, where with more antennas at the relay, the performance will be better. }

\subsection{High Relay Power and Low Source Powers}

\begin{figure}[t]
\begin{centering}
\includegraphics[scale=.40]{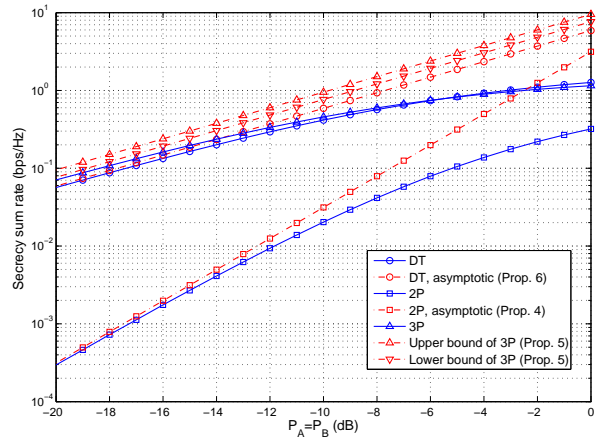}
\vspace{-0.1cm}
\centering
 \caption{Comparison of the three schemes with high relay power when $N_A=2$, $N_R=3$, $N_B=2$ and $P_R =40$ dB.}\label{fig:P_R_infty_P_i_0}
\end{centering}
\vspace{-0.3cm}
\end{figure}

Fig. \ref{fig:P_R_infty_P_i_0} shows the performance of three schemes when $P_R = 40$dB and the source powers are low.  We find that the two-phase scheme is much worse than the other schemes and three-phase scheme is better than the direct transmission scheme, which verifies Corollary \ref{corollary_P_R_infty_P_i_0}. By careful observation, we see that $R^{2P}_{\max}$ decreases to zero as twice faster as $R^{DT}_{\max}$ and $R^{3P}_{\max}$ when the source powers tend to zero. Moreover, we also find that the asymptotical results are quite accurate when the source powers are low.

\subsection{Low Relay Power}

\begin{figure}[t]
\begin{centering}
\includegraphics[scale=.40]{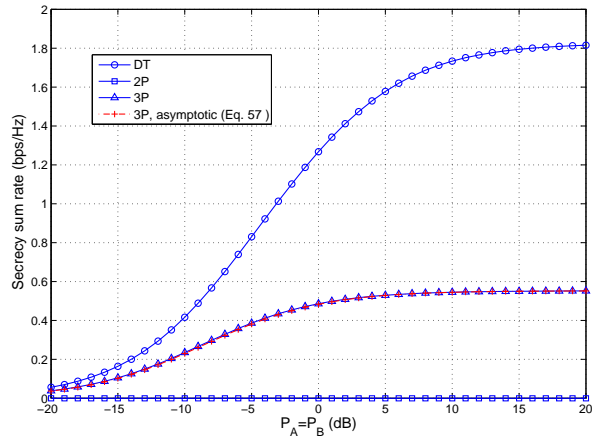}
\vspace{-0.1cm}
\centering
 \caption{Comparison of the three schemes with low relay power. $N_A=2$, $N_R=3$, $N_B=2$ and $P_R =-20$ dB.}\label{fig:Low_relay_power}
\end{centering}
\vspace{-0.3cm}
\end{figure}

In Fig. \ref{fig:Low_relay_power}, we compare the three schemes when the relay power is as low as -20dB. We find that the maximum secrecy sum rate of two-phase scheme is close to zero and direct transmission is better than three-phase scheme, which verifies Corollary \ref{corollary_P_R_0}. The reason is that the only link $A\leftrightarrows R \leftrightarrows B$ of the two-phase scheme is very weak while there are strong direct links in the other two schemes with high source powers.


\subsection{General Relay SNR and Fading Channels}
\begin{figure}[t]
\begin{centering}
\includegraphics[scale=.40]{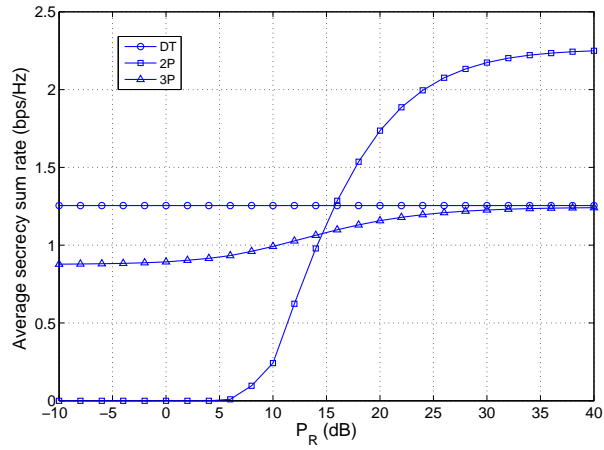}
\vspace{-0.1cm}
\centering
 \caption{Comparison of the three schemes with varying relay power, $P_A=P_B=15$ dB, $N_A=N_B=2$, $N_R=3$.}\label{fig:Rate_vs_P_R}
\end{centering}
\vspace{-0.3cm}
\end{figure}

\begin{figure}[t]
\begin{centering}
\includegraphics[scale=.40]{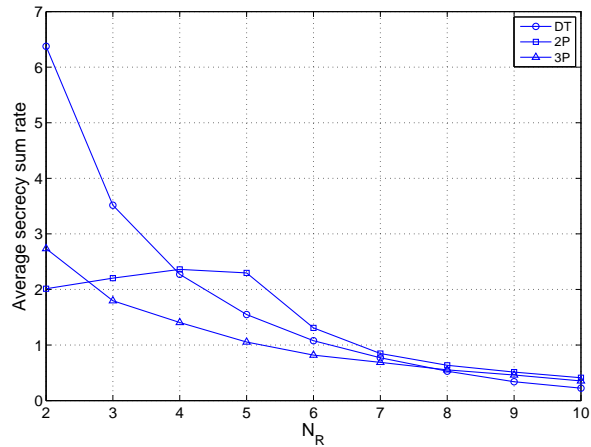}
\vspace{-0.1cm}
\centering
 \caption{Comparison of the three schemes with varying relay antenna number, $P_A=P_B=15$dB, $P_R=25$dB.}\label{fig:Rate_vs_N_R}
\end{centering}
\vspace{-0.3cm}
\end{figure}

{\color{black}{We have considered the high relay power and low relay power case. In this subsection, we consider the general relay power. For this case, this is no asymptotic results and we perform simulation with $1000$ different channel realizations (every entry of the matrices is generated from $\mathcal{CN}(0,1)$ distribution) and obtain average secrecy sum rate. For the two-phase and three-phase scheme, the simulation results are obtained by Algorithm 1 and 2.

In Fig. \ref{fig:Rate_vs_P_R}, we compare the average secrecy sum rates of the three schemes with varying relay power. The source powers are fixed at $15$dB and $N_A=N_B=2$, $N_R=3$. The average rate of the direct transmission scheme does not change with the relay power as the relay does not transmit in this scheme. The average rate of the three-phase scheme increases with the relay power and has similar performance with direct transmission at high relay power. For the two-phase scheme, as the relay power increases, the average rate rises from zero to as high as $2.2$ bps/Hz.  We can see that the two-phase scheme is much better than the other two schemes when relay power is high. The reason is that in this case, signal alignment can be achieved when $P_R$ is large as $N_A+N_B>N_R$.

In Fig. \ref{fig:Rate_vs_N_R}, we plot the average secrecy sum rate versus the relay antenna number. The source node $A$ and $B$ both have three antennas. The relay power is $25$dB and the source powers are $15$dB. From the figure, we see that the average rate of the direct transmission scheme monotonically decreases with $N_R$. The reason is that the untrusted relay will be more powerful to eavesdrop the direct transmission signal as $N_R$ increases. For the two-phase transmission scheme, the average rate achieves the largest value when $N_R =4$. The reason is that when $N_R$ is small, the relay does not have enough abilities to help the two-way transmission and when $N_R$ is large, the relay will be more powerful to decode the received signals. For the three-phase scheme, the average rate also decreases with $N_R$ in this case.
}}

\section{Conclusion}

In this paper, we investigated a MIMO two-way AF relay system where the two source nodes exchange confidential information with an untrusted relay. For both two-phase and three-phase two-way relay schemes, we proposed efficient algorithms to jointly design the secure source and relay beamformers iteratively. Furthermore, we analyzed the asymptotical performance of the secure beamforming schemes in low and high power regimes of the sources and relay. Simulation results validate our asymptotical analysis.

From these results, we can conclude that the conventional two-way direct transmission is preferred when the relay power goes to zero. When the relay power approaches infinity and source powers approach zero, the three-phase two-way relay scheme performs best. Moreover, when all powers go to infinity, the two-phase two-way relay scheme has the best performance if signal alignment techniques are used, which also lowers the requirement of numbers of antennas at the source nodes for security.

\appendices
\section{Secure Beamforming of Two-Way Direct Transmission Scheme}\label{app_DT}
For the two-way direct transmission scheme, the transmission consists of two time slots. In the first time slot, $A$ transmits while $B$ and $R$ listen. During the second time slot, $B$ transmits while $A$ and $R$ listens. The received signals at $B$ and $R$ in the first time slot are respectively given by
\begin{eqnarray}
  \mb{y}_B^{DT} &=& \mb{T}_A \mb{q}_A s_A + \mb{n}_B, \\
  \mb{y}_{R1}^{DT} &=& \mb{H}_A \mb{q}_A s_A + \mb{n}_{R1},
\end{eqnarray}
and the received signals in the second time slot are similar.

An achievable secrecy sum rate of this two-way direct transmission scheme given by \cite{Khisti_IT10a} is,
\begin{equation}
    \mathcal{R}_s^{DT} = \sum\limits_{i \in \left\{ {A,B} \right\}} {\frac{1}{2}{{\left[ {{{\log }_2}\frac{{1 + {\mb{q}}_i^H{\mb{T}}_i^H{{\mb{T}}_i}{{\mb{q}}_i}}}{{1 + {\mb{q}}_i^H{\mb{H}}_i^H{{\mb{H}}_i}{{\mb{q}}_i}}}} \right]}^ + }}.
\end{equation}

We want to maximize the secrecy sum rate $\mathcal{R}_s^{DT}$ subject to the source power constraints. According to \cite{Shafiee_ISIT07}, \cite{Khisti_IT10a} and \cite{Jeong_TSP12}, the optimal beamforming $\mb{q}_i^{DT}$ of the two-way direct transmission scheme is given by
\begin{equation}\label{dt_Beamforming}
  \mb{q}_i^{DT}= \frac{\sqrt{P_i^{DT}}\pmb{\psi}_{max}({\mb{I}} + {P_i^{DT}}{\mb{T}}_i^H{{\mb{T}}_i}, {\mb{I}} + {P_i^{DT}}{\mb{H}}_i^H{{\mb{H}}_i})}{\| \pmb{\psi}_{max}({\mb{I}} + {P_i^{DT}}{\mb{T}}_i^H{{\mb{T}}_i}, {\mb{I}} + {P_i^{DT}}{\mb{H}}_i^H{{\mb{H}}_i}) \|},~ i\in \{A, B\},
\end{equation}
and the maximum secrecy sum rate is given by
\begin{eqnarray}\label{nc_rate}
  & &\mathcal{R}_{\max}^{DT}(P_A^{DT}, P_B^{DT}) \nonumber \\
  &=& \sum\limits_{i \in \left\{ {A,B} \right\}} {\frac{1}{2}{{\left[ {\log_2 {\lambda _{\max }}\left( {{\mb{I}} + {P_i^{DT}}{\mb{T}}_i^H{{\mb{T}}_i},{\mb{I}} + {P_i^{DT}}{\mb{H}}_i^H{{\mb{H}}_i}} \right)} \right]}^ + }},
\end{eqnarray}
where the factor of $\frac{1}{2}$ is due to the use of two orthogonal phases.

\section{Search $\mb{A}$ using Gradient Method}
\label{app_2P_subproblem_A}
Substituting \eqref{eq_2P_structure} into \eqref{eq_2P_problem}, we obtain a subproblem of optimizing $\mb{A}$ given $\mb{q}_A$ and $\mb{q}_B$ as follows,
\begin{subequations}\label{eq_2P_subproblem_A}
\begin{align}
  &\mathop{\min}\limits_{\mb{A}} & & -\mathcal{R}_s^{2P} \\
  &\text{s. t.} & &\mathrm{Tr}\big(\mb{A} \mb{U}^H \mb{H}_A \mb{q}_A \mb{q}_A^H \mb{H}_A^H \mb{U} \mb{A}^H  \\
  & & &+ \mb{A} \mb{U}^H \mb{H}_B \mb{q}_B \mb{q}_B^H \mb{H}_B^H \mb{U} \mb{A}^H + \mb{A} \mb{A}^H　\big) \leq P_R^{2P}. \label{eq_2P_subproblem_A_cons}
  \end{align}
\end{subequations}

The logarithmic barrier function associated with \eqref{eq_2P_subproblem_A} is,
\begin{eqnarray}
B(\mb{A}, \mu) &=& -\mathcal{R}_s^{2P} - \mu \ln \Big(P_R^{2P}-\mathrm{Tr}\big(\mb{A} \mb{U}^H \mb{H}_A \mb{q}_A \mb{q}_A^H \mb{H}_A^H \mb{U} \mb{A}^H \nonumber \\
& &+ \mb{A} \mb{U}^H \mb{H}_B \mb{q}_B \mb{q}_B^H \mb{H}_B^H \mb{U} \mb{A}^H + \mb{A} \mb{A}^H　\big)\Big)
\end{eqnarray}
where $\mu>0$ is the barrier parameter.

The gradient of $B(\mb{A}, \mu)$ with respect to $\mb{A}$ is given by \eqref{eq_gradient} shown at the top of the next page,
\begin{figure*}[t]
\begin{eqnarray}\label{eq_gradient}
\frac{{\partial B(\mb{A}, \mu)}}{{\partial {{\mb{A}}^*}}} &=& - \sum\limits_{i \in \left\{ {A,B} \right\}} \log_2 e {\frac{{{\mb{V}^H}{\mb{G}}_i^H{\mb{K}}_i^{-1}{{\mb{G}}_i}{\mb{F}}{{\mb{H}}_{\bar{i}}}{{\mb{q}}_{\bar{i}}}{\mb{q}}_{\bar{i}}^H{\mb{H}}_{\bar{i}}^H {\mb{U}} - {\mb{V}^H} {\mb{G}}_i^H{\mb{K}}_i^{ - 1}{{\mb{G}}_i}{\mb{F}}{{\mb{H}}_{\bar{i}}}{{\mb{q}}_{\bar{i}}}{\mb{q}}_{\bar{i}}^H{\mb{H}}_{\bar{i}}^H{{\mb{F}}^H}{\mb{G}}_i^H{\mb{K}}_i^{ - 1}{{\mb{G}}_i}{\mb{F}}{\mb{U}}}}
{{2{{\left( {1 + {\mb{q}}_{\bar{i}}^H{\mb{H}}_{\bar{i}}^H{\mb{F}}^H{\mb{G}}_i^H{\mb{K}}_i^{ - 1}{{\mb{G}}_i} {\mb{F}} {{\mb{H}}_{\bar{i}}}{{\mb{q}}_{\bar{i}}}} \right)}}}}}  \nonumber \\
& & + \mu \frac{\mb{A} \mb{U}^H \mb{H}_A \mb{q}_A \mb{q}_A^H \mb{H}_A^H \mb{U} + \mb{A} \mb{U}^H \mb{H}_B \mb{q}_B \mb{q}_B^H \mb{H}_B^H \mb{U} + \mb{A} }{{{P_R^{2P}} - \mathrm{Tr}\big(\mb{A} \mb{U}^H \mb{H}_A \mb{q}_A \mb{q}_A^H \mb{H}_A^H \mb{U} \mb{A}^H + \mb{A} \mb{U}^H \mb{H}_B \mb{q}_B \mb{q}_B^H \mb{H}_B^H \mb{U} \mb{A}^H + \mb{A} \mb{A}^H　\big)}}
\end{eqnarray}
\hrulefill
\end{figure*}
With this gradient, we use gradient descent method to search $\mb{A}$.

\section{Search Optimal $\mb{q}_B$ Given $\mb{F}$ and $\mb{q}_A$ in two-phase two-way relay Scheme}
\label{app_2P_q_B_design}
First, we rewrite \eqref{eq_2P_problem} in the homogenized form with respect to $\mb{q}_B$, as \eqref{eq_homogenized_form} shown at the top of the next page.
\begin{figure*}[t]
\begin{subequations}\label{eq_homogenized_form}
\begin{align}
& \mathop {\max }\limits_{{{\mb{q}}_B},t} & &\frac{{\mathrm{Tr}\left\{ {\left[ {\begin{array}{*{20}{c}}
{{\mb{H}}_B^H{{\mb{F}}^H}{\mb{G}}_A^H{\mb{K}}_A^{ - 1}{{\mb{G}}_A}{\mb{F}}{{\mb{H}}_B}}&{\mb{0}}\\
{\mb{0}}&1
\end{array}} \right]\left[ {\begin{array}{*{20}{c}}
{{{\mb{q}}_B}{\mb{q}}_B^H}&{{{\mb{q}}_B}{t^*}}\\
{{\mb{q}}_B^Ht}&{{{\left| t \right|}^2}}
\end{array}} \right]} \right\}}}{{\mathrm{Tr}\left\{ {\left[ {\begin{array}{*{20}{c}}
{{\mb{H}}_B^H\left( {\left( {1 + {{\left\| {{{\mb{H}}_A}{{\mb{q}}_A}} \right\|}^2}} \right){\mb{I}} - {{\mb{H}}_A}{{\mb{q}}_A}{\mb{q}}_A^H{\mb{H}}_A^H} \right){{\mb{H}}_B}}&{\mb{0}}\\
{\mb{0}}&{1 + {{\left\| {{{\mb{H}}_A}{{\mb{q}}_A}} \right\|}^2}}
\end{array}} \right]\left[ {\begin{array}{*{20}{c}}
{{{\mb{q}}_B}{\mb{q}}_B^H}&{{{\mb{q}}_B}{t^*}}\\
{{\mb{q}}_B^Ht}&{{{\left| t \right|}^2}}
\end{array}} \right]} \right\}}} \\
& s.t. & & \mathrm{Tr}\left\{ {\left[ {\begin{array}{*{20}{c}}
{\mb{I}}&{\mb{0}}\\
{\mb{0}}&0
\end{array}} \right]\left[ {\begin{array}{*{20}{c}}
{{{\mb{q}}_B}{\mb{q}}_B^H}&{{{\mb{q}}_B}{t^*}}\\
{{\mb{q}}_B^Ht}&{{{\left| t \right|}^2}}
\end{array}} \right]} \right\} \le {P_B},  \\
& & &{\rm{Tr}}\left\{ {\left[ {\begin{array}{*{20}{c}}
{{\mb{H}}_B^H{{\mb{F}}^H}{\mb{F}}{{\mb{H}}_B}}&{\mb{0}}\\
{\mb{0}}&0
\end{array}} \right]\left[ {\begin{array}{*{20}{c}}
{{{\mb{q}}_B}{\mb{q}}_B^H}&{{{\mb{q}}_B}{t^*}}\\
{{\mb{q}}_B^Ht}&{{{\left| t \right|}^2}}
\end{array}} \right]} \right\} \le {P_r} - \rm{Tr}\left\{ {{\mb{F}}{{\mb{F}}^H}} \right\} - \rm{Tr}\left\{ {{\mb{F}}{{\mb{H}}_A}{{\mb{q}}_A}{\mb{q}}_A^H{\mb{H}}_A^H{{\mb{F}}^H}} \right\}, \\
& & &{\rm{Tr}}\left\{ {\left[ {\begin{array}{*{20}{c}}
{\mb{0}}&{\mb{0}}\\
{\mb{0}}&1
\end{array}} \right]\left[ {\begin{array}{*{20}{c}}
{{{\mb{q}}_B}{\mb{q}}_B^H}&{{{\mb{q}}_B}{t^*}}\\
{{\mb{q}}_B^Ht}&{{{\left| t \right|}^2}}
\end{array}} \right]} \right\} = 1.
\end{align}
\end{subequations}
\hrulefill
\end{figure*}
 Then, we can follow the same procedure in \cite[Section III-B]{Maio_TSP11} or \cite[Appendix A]{Jeong_TSP12} to find the optimal $\mb{q}_B$. The basic idea is to first relax \eqref{eq_homogenized_form} into a fractional semidefinite programming problem, which is then transformed to a SDP problem using Charnes-Cooper variable transformation. At last, the rank-one matrix decomposition theorem \cite[Theorem 2.3]{Ai_Math11} is used. Here we omit the details.

\section{Proof of Lemma \ref{lemma_3P_structure}}
\label{app_3P_structure}
First, we consider the case where $N_R>N_A+N_B$.

Without loss of generality, we can express $\mb{F}_i$ as
\begin{equation}
\label{eq_app_3P_F}
{{\mb{F}}_i} = \left[ {\begin{array}{*{20}{c}}
{{{\mb{V}}}}&{{\mb{V}}^ \bot }
\end{array}} \right]\left[ {\begin{array}{*{20}{c}}
{{{\mb{a}}_i}}&{{{\mb{B}}_i}}\\
{{{\mb{c}}_i}}&{{{\mb{D}}_i}}
\end{array}} \right]\left[ {\begin{array}{*{20}{c}}
{{\mb{U}}_i^H}\\
{{\mb{U}}_i^{ \bot H}}
\end{array}} \right]
\end{equation}
where $\mb{V}$ is from \eqref{eq_qr_3P} , $\mb{V}^ \bot \in \mathbb{C}^{N_R \times (N_R - N_A -N_B)}$ such that $\left[\mb{V} \quad \mb{V}^{\perp}\right]$ is unitary , $\mb{U}_i$ is $\frac{{{\mb{H}}_i{\mb{q}}_i}}{{\left\| {{{\mb{H}}_i}{{\mb{q}}_i}} \right\|}}$ , $\mb{U}_i^ \bot \in \mathbb{C}^{N_R \times (N_R-1)}$ such that $\left[\mb{U}_i \quad \mb{U}_i^{\perp}\right]$ is unitary, and $\mb{a}_i \in \mathbb{C}^{(N_A+ N_B) \times 1}$, $\mb{c}_i \in \mathbb{C}^{(N_R - N_A- N_B) \times 1}$, $\mb{B}_i \in \mathbb{C}^{(N_A+ N_B) \times (N_R-1)}$, $\mb{D}_i \in \mathbb{C}^{(N_R -N_A -N_B) \times (N_R -1)}$.
Therefore, we obtain \eqref{eq_x_BA} shown at the top of the next page.
\begin{figure*}[t]
\begin{eqnarray}\label{eq_x_BA}
    x_{BA}&\triangleq &{\mb{q}}_B^H{\mb{H}}_B^H{\mb{F}}_B^H{\mb{G}}_A^H{\left( {{{\mb{G}}_A}\left( {{{\mb{F}}_A}{\mb{F}}_A^H + {{\mb{F}}_B}{\mb{F}}_B^H} \right){\mb{G}}_A^H + {\mb{I}}} \right)^{ - 1}}{{\mb{G}}_A}{{\mb{F}}_B}{{\mb{H}}_B}{{\mb{q}}_B} \nonumber \\
  &\stackrel{(a)}{=}& {{{{\left\| {{{\mb{H}}_B}{{\mb{q}}_B}} \right\|}^2}}}{\mb{a}}_2^H{\mb{V}}^H{\mb{G}}_A^H{\left( {\sum\limits_{i = 1}^2 {{{\mb{G}}_A}{{\mb{V}}}{{\mb{a}}_i}{\mb{a}}_i^H{\mb{V}}^H{\mb{G}}_A^H}  + \sum\limits_{i = 1}^2 {{{\mb{G}}_A}{{\mb{V}}}{{\mb{B}}_i}{\mb{B}}_i^H{\mb{V}}^H{\mb{G}}_A^H}  + {\mb{I}}} \right)^{ - 1}}{{\mb{G}}_A}{{\mb{V}}}{{\mb{a}}_2} \nonumber \\
  &\stackrel{(b)}{\leq}& {{{{\left\| {{{\mb{H}}_B}{{\mb{q}}_B}} \right\|}^2}}}{\mb{a}}_2^H{\mb{V}}^H{\mb{G}}_A^H{\left( {\sum\limits_{i = 1}^2 {{{\mb{G}}_A}{{\mb{V}}}{{\mb{a}}_i}{\mb{a}}_i^H{\mb{V}}^H{\mb{G}}_A^H} + {\mb{I}}} \right)^{ - 1}}{{\mb{G}}_A}{{\mb{V}}}{{\mb{a}}_2}
\end{eqnarray}
\hrulefill
\end{figure*}
Therein, $(a)$ is from the above property of $\mb{F}_i$ \eqref{eq_app_3P_F}, $(b)$ is from that ${\sum\limits_{i = 1}^2 {{{\mb{G}}_A}{{\mb{U}}_i}{{\mb{B}}_i}{\mb{B}}_i^H{\mb{U}}_i^H{\mb{G}}_A^H} }$ is positive semidefinite matrix.
We see that the information rate from B to A $\mathcal{R}_{BA}^{3P}=\log_2 (1+ \mb{q}_B^H \mb{H}_B^H \mb{H}_B \mb{q}_B +  x_{BA})$ is not related to $\mb{c}_i$ and $\mb{D}_i$ and achieves a upper bound when $\mb{B}_i = \mb{0}$. Similarly, the information rate from A to B, $\mathcal{R}_{AB}^{3P}$, is also not related to $\mb{c}_i$ and $\mb{D}_i$ and achieves a upper bound when $\mb{B}_i = \mb{0}$. In addition, the power consumed by the relay is
\begin{eqnarray*}
   & &\mathrm{Tr}\big(\mb{F}_A \mb{H}_A \mb{q}_A \mb{q}_A^H \mb{H}_A^H \mb{F}_A^H + \mb{F}_B \mb{H}_B \mb{q}_B \mb{q}_B^H \mb{H}_B^H \mb{F}_B^H \\
   & &\quad \quad \quad + \mb{F}_A \mb{F}_A^H + \mb{F}_B \mb{F}_B^H　\big) \\
   &=&
  {\left\| {{{\mb{H}}_A}{{\mb{q}}_A}} \right\|^2}\left( {{{\left\| {{{\mb{a}}_1}} \right\|}^2} + {{\left\| {{{\mb{c}}_1}} \right\|}^2}} \right) + {\left\| {{{\mb{H}}_B}{{\mb{q}}_B}} \right\|^2}\left( {{{\left\| {{{\mb{a}}_2}} \right\|}^2} + {{\left\| {{{\mb{c}}_2}} \right\|}^2}} \right){\rm{ }} \\
  & &+ \sum\limits_{i = 1}^2 {{{\left\| {{{\mb{a}}_i}} \right\|}^2}}  + \sum\limits_{i = 1}^2 {\left\| {{{\mb{B}}_i}} \right\|_F^2}  + \sum\limits_{i = 1}^2 {{{\left\| {{{\mb{c}}_i}} \right\|}^2}}  + \sum\limits_{i = 1}^2 {\left\| {{{\mb{D}}_i}} \right\|_F^2}
 \end{eqnarray*}
 We find that the relay power is increased when $\mb{B}_i$, $\mb{c}_i$, $\mb{D}_i$ is not zero.
 Therefore, it leads to $\mb{B}_i=\mb{0}$, $\mb{c}_i=\mb{0}$ and $\mb{D}_i=\mb{0}$.

 When $N_R \leq N_A+N_B$, we can express $\mb{F}_i$ as
\begin{equation}
\label{eq_app_3P_F_second}
{{\mb{F}}_i} = {\mb{V}}\left[ {\begin{array}{*{20}{c}}
  {{{\mb{a}}_{\mb{i}}}}&{{{\mb{B}}_{\mb{i}}}}
\end{array}} \right]\left[ {\begin{array}{*{20}{c}}
  {{\mb{U}}_{\mb{i}}^{\mb{H}}} \\
  {{\mb{U}}_{\mb{i}}^{ \bot {\mb{H}}}}
\end{array}} \right]
\end{equation}
where $\mb{V}$ is from \eqref{eq_qr_3P} , $\mb{U}_i$ is $\frac{{{\mb{H}}_i {\mb{q}}_i}}{{\left\| {{{\mb{H}}_i}{{\mb{q}}_i}} \right\|}}$ , $\mb{U}_i^ \bot \in \mathbb{C}^{N_R \times (N_R-1)}$ such that $\left[\mb{U}_i \quad \mb{U}_i^{\perp}\right]$ is unitary, and $\mb{a}_i \in \mathbb{C}^{N_R \times 1}$, $\mb{B}_i \in \mathbb{C}^{N_R \times (N_R-1)}$. Similar as the above case, we can prove that the optimal $\mb{B}_i=\mb{0}$.

\section{Proof of Proposition \ref{proposition_2P_P_R_infty_P_i_0}}
\label{app_2P_P_R_infty_P_i_0}

Plugging the condition $P_A \to 0$, $P_B \to 0$ into \eqref{eq_Rate_2P_P_R_infty}, we have
\begin{eqnarray*}
& &\lim \limits_{P_R \to \infty} \mathcal{R}_s^{2P} \\
&=&\frac{1}{2}{\log _2}\frac{1}{{1 - \frac{{{\mb{q}}_B^H{\mb{H}}_B^H{{\mb{H}}_A}{{\mb{q}}_A}{\mb{q}}_A^H{\mb{H}}_A^H{{\mb{H}}_B}{{\mb{q}}_B}}}{{\left( {1 + {\mb{q}}_B^H{\mb{H}}_B^H{{\mb{H}}_B}{{\mb{q}}_B}} \right)\left( {1 + {\mb{q}}_A^H{\mb{H}}_A^H{{\mb{H}}_A}{{\mb{q}}_A}} \right)}}}}\\
 &=&  - \frac{1}{2}{\log _2}\left( {1 - \frac{{{\mb{q}}_B^H{\mb{H}}_B^H{{\mb{H}}_A}{{\mb{q}}_A}{\mb{q}}_A^H{\mb{H}}_A^H{{\mb{H}}_B}{{\mb{q}}_B}}}{{\left( {1 + {\mb{q}}_B^H{\mb{H}}_B^H{{\mb{H}}_B}{{\mb{q}}_B}} \right)\left( {1 + {\mb{q}}_A^H{\mb{H}}_A^H{{\mb{H}}_A}{{\mb{q}}_A}} \right)}}} \right)\\
 &\approx &  - \frac{1}{2}{\log _2}\left( {1 - {\mb{q}}_B^H{\mb{H}}_B^H{{\mb{H}}_A}{{\mb{q}}_A}{\mb{q}}_A^H{\mb{H}}_A^H{{\mb{H}}_B}{{\mb{q}}_B}} \right)\\
 &\approx & \frac{1}{{2\ln 2}}{\left\| {{\mb{q}}_B^H{\mb{H}}_B^H{{\mb{H}}_A}{{\mb{q}}_A}} \right\|^2}.
\end{eqnarray*}
To maximize $\|\mb{q}_B^H \mb{H}_B^H \mb{H}_A \mb{q}_A \|^2$, we obtain Proposition \ref{proposition_2P_P_R_infty_P_i_0}.

\section{Proof of Proposition \ref{proposition_3P_P_R_infty_P_i_infty}}
\label{app_3P_P_R_infty_P_i_infty}
Substituting the optimal relay beamforming structure \eqref{eq_3P_structure} into \eqref{eq_3P_B_A_rate}, we obtain the third term in \eqref{eq_3P_B_A_rate} as follows,
\small
\begin{eqnarray*}
 & &{\mb{q}}_B^H{\mb{H}}_B^H{\mb{F}}_B^H{\mb{G}}_A^H{{\left( {{{\mb{G}}_A}\left( {{{\mb{F}}_A}{\mb{F}}_A^H + {{\mb{F}}_B}{\mb{F}}_B^H} \right){\mb{G}}_A^H + {\mb{I}}} \right)}^{ - 1}}{{\mb{G}}_A}{{\mb{F}}_B}{{\mb{H}}_B}{{\mb{q}}_B} \\
&=&{{\left\| {{{\mb{H}}_B}{{\mb{q}}_B}} \right\|}^2}{\mb{a}}_2^H{\mb{V}}^H{\mb{G}}_A^H{\left( {\sum\limits_{i = 1}^2 {{{\mb{G}}_A}{{\mb{V}}}{{\mb{a}}_i}{\mb{a}}_i^H{\mb{V}}^H{\mb{G}}_A^H} + {\mb{I}}} \right)^{ - 1}}{{\mb{G}}_A}{{\mb{V}}}{{\mb{a}}_2} \\
 &\stackrel{(a)}\le & {\left\| {{{\mb{H}}_B}{{\mb{q}}_B}} \right\|^2}{\mb{a}}_2^H{{\mb{V}}^H}{\mb{G}}_A^H{\left( {{{\mb{G}}_A}{\mb{V}}{{\mb{a}}_2}{\mb{a}}_2^H{{\mb{V}}^H}{\mb{G}}_A^H + {\mb{I}}} \right)^{ - 1}}{{\mb{G}}_A}{\mb{V}}{{\mb{a}}_2}\\
 &\stackrel{(b)}=& {\left\| {{{\mb{H}}_B}{{\mb{q}}_B}} \right\|^2}{\mb{a}}_2^H{{\mb{V}}^H}{\mb{G}}_A^H\Big( {\mb{I}} - {{\mb{G}}_A}{\mb{V}}{{\mb{a}}_2}{{\left( {{\mb{a}}_2^H{{\mb{V}}^H}{\mb{G}}_A^H{{\mb{G}}_A}{\mb{V}}{{\mb{a}}_2} + 1} \right)}^{ - 1}}\cdot \\
 & &{\mb{a}}_2^H{{\mb{V}}^H}{\mb{G}}_A^H \Big){{\mb{G}}_A}{\mb{V}}{{\mb{a}}_2}\\
 &=& {\left\| {{{\mb{H}}_B}{{\mb{q}}_B}} \right\|^2}\frac{{{{\left\| {{{\mb{G}}_A}{\mb{V}}{{\mb{a}}_2}} \right\|}^2}}}{{1 + {{\left\| {{{\mb{G}}_A}{\mb{V}}{{\mb{a}}_2}} \right\|}^2}}} \\
 &\leq & {\left\| {{{\mb{H}}_B}{{\mb{q}}_B}} \right\|^2}
\end{eqnarray*}
\small
\normalsize
where $(a)$ is from that ${{{\mb{G}}_A}{\mb{V}}{{\mb{a}}_1}{\mb{a}}_1^H{{\mb{V}}^H}{\mb{G}}_A^H}$ is positive semidefinite, $(b)$ is from the matrix inverse lemma.

The above third term in \eqref{eq_3P_B_A_rate} also has a lower bound by simply letting $\mb{a}_1 = \mb{a}_2 = \mb{\bar{a}}$,
\begin{eqnarray*}
  & &{\mb{q}}_B^H{\mb{H}}_B^H{\mb{F}}_B^H{\mb{G}}_A^H{{\left( {{{\mb{G}}_A}\left( {{{\mb{F}}_A}{\mb{F}}_A^H + {{\mb{F}}_B}{\mb{F}}_B^H} \right){\mb{G}}_A^H + {\mb{I}}} \right)}^{ - 1}} \cdot\\
 & & {{\mb{G}}_A}{{\mb{F}}_B}{{\mb{H}}_B}{{\mb{q}}_B}\\
 &=& {\left\| {{{\mb{H}}_B}{{\mb{q}}_B}} \right\|^2}{{\mb{a}}^H}{{\mb{V}}^H}{\mb{G}}_A^H{\left( {2{{\mb{G}}_A}{\mb{Va}}{{\mb{a}}^H}{{\mb{V}}^H}{\mb{G}}_A^H + {\mb{I}}} \right)^{ - 1}}{{\mb{G}}_A}{\mb{Va}}\\
&=&\frac{1}{2}{\left\| {{{\mb{H}}_B}{{\mb{q}}_B}} \right\|^2}\left( {1 - {{\left( {1 + 2{{\mb{a}}^H}{{\mb{V}}^H}{\mb{G}}_A^H{{\mb{G}}_A}{\mb{Va}}} \right)}^{ - 1}}} \right)\\
&=&\frac{1}{2}{\left\| {{{\mb{H}}_B}{{\mb{q}}_B}} \right\|^2}\frac{{2{{\left\| {{{\mb{G}}_A}{\mb{Va}}} \right\|}^2}}}{{1 + 2{{\left\| {{{\mb{G}}_A}{\mb{Va}}} \right\|}^2}}} \\
&\rightarrow & \frac{1}{2}{\left\| {{{\mb{H}}_B}{{\mb{q}}_B}} \right\|^2} \quad \text{as} \quad P_R\to \infty
 \end{eqnarray*}

 Therefore, we have
 \begin{eqnarray} \label{eq_3P_UB_LB_rate}
   \frac{1}{3}{\log _2}\left( {1 + {\mb{q}}_i^H{\mb{T}}_i^H{{\mb{T}}_i}{{\mb{q}}_i} + \frac{1}{2}{\mb{q}}_i^H{\mb{H}}_i^H{{\mb{H}}_i}{{\mb{q}}_i}} \right) \le \mathop {\lim }\limits_{{P_R} \to \infty } \mathcal{R}_{i\bar{i}}^{3P} \nonumber \\
   \le \frac{1}{3}{\log _2}\left( {1 + {\mb{q}}_i^H{\mb{T}}_i^H{{\mb{T}}_i}{{\mb{q}}_i} + {\mb{q}}_i^H{\mb{H}}_i^H{{\mb{H}}_i}{{\mb{q}}_i}} \right).
 \end{eqnarray}

To prove Proposition \ref{proposition_3P_P_R_infty_P_i_infty}, we first substitute the upper bound and lower bound into \eqref{eq_3P_secrecy_sum_rate}. After that, the proof procedure of Proposition \ref{proposition_3P_P_R_infty_P_i_infty} is similar to the proof of \cite[Lemma 7]{Jeong_TSP12}. In addition, we assume that the entries of channel matrices are generated from continuous distribution.

\section{Proof of Corollary \ref{corollary_P_R_0}}
\label{app_proposition_P_R_0}
For fair comparison, we set $P_i = P_i^{DT} = P_i^{2P} = \frac{2}{3}P_i^{3P}$, $i \in \{A,B\}$ and $P_R = P_R^{2P} = \frac{2}{3}P_R^{3P} $. When the relay power $P_R\to 0$, there are only direct links between the two source nodes for the three-phase scheme. Thus, the maximum secrecy sum rate of the three-phase two-way relay scheme $\mathcal{R}_{\max}^{3P}$ is
\begin{eqnarray}
     & &\mathcal{R}_{\max}^{3P}  \\
     &\approx & \mathop{\max} \limits_{\mb{q}_A, \mb{q}_B} \frac{1}{3} \sum\limits_{i \in \left\{ {A,B} \right\}} {{{\left[ {{{\log }_2}\frac{{1 + {\mb{q}}_i^H{\mb{T}}_i^H{{\mb{T}}_i}{{\mb{q}}_i}}}{{1 + {\mb{q}}_i^H{\mb{H}}_i^H{{\mb{H}}_i}{{\mb{q}}_i}}}} \right]}^ + }}  \nonumber \\
    &=& \frac{1}{3} \sum\limits_{i \in \left\{ {A,B} \right\}} \left[ \log_2 \left( {{\lambda _{\max }}\left( {{\mb{I}} + {P_i^{3P}}{\mb{T}}_i^H{{\mb{T}}_i},{\mb{I}} + {P_i^{3P}}{\mb{H}}_i^H{{\mb{H}}_i}} \right)} \right) \right]^+ \nonumber\\
    &=& \frac{1}{3} \sum\limits_{i \in \left\{ {A,B} \right\}} \left[ \log_2 \left( {{\lambda _{\max }}\left( {{\mb{I}} + \frac{3}{2}{P_i}{\mb{T}}_i^H{{\mb{T}}_i},{\mb{I}} + \frac{3}{2}{P_i}{\mb{H}}_i^H{{\mb{H}}_i}} \right)} \right) \right]^+. \nonumber
\end{eqnarray}
In addition, we have
    \begin{eqnarray*}
  & &{\lambda _{\max }}\left( {{\mb{I}} + \frac{3}{2}{P_i}{\mb{T}}_i^H{{\mb{T}}_i},{\mb{I}} + \frac{3}{2}{P_i}{\mb{H}}_i^H{{\mb{H}}_i}} \right)  \\
   &\stackrel{(a)}=&{\lambda _{\max }}\left( {\frac{3}{2}{P_i}{\mb{T}}_i^H{{\mb{T}}_i} - \frac{3}{2}{P_i}{\mb{H}}_i^H{{\mb{H}}_i},{\mb{I}} + \frac{3}{2}{P_i}{\mb{H}}_i^H{{\mb{H}}_i}} \right) + 1  \\
   &=& \mathop {\max }\limits_{\pmb{\psi}}  \frac{{{{\mb{\pmb{\psi} }}^H}\left( {\frac{3}{2}{P_i}{\mb{T}}_i^H{{\mb{T}}_i} - \frac{3}{2}{P_i}{\mb{H}}_i^H{{\mb{H}}_i}} \right){\mb{\pmb{\psi} }}}}{{{{\mb{\pmb{\psi} }}^H}\left( {{\mb{I}} + \frac{3}{2}{P_i}{\mb{H}}_i^H{{\mb{H}}_i}} \right){\mb{\pmb{\psi} }}}} + 1  \\
   &\leq& \mathop {\max } \limits_{\pmb{\psi}}  \frac{3}{2}\frac{{{{\mb{\pmb{\psi} }}^H}\left( {{P_i}{\mb{T}}_i^H{{\mb{T}}_i} - {P_i}{\mb{H}}_i^H{{\mb{H}}_i}} \right){\mb{\pmb{\psi} }}}}{{{{\mb{\pmb{\psi} }}^H}\left( {{\mb{I}} + {P_i}{\mb{H}}_i^H{{\mb{H}}_i}} \right){\mb{\pmb{\psi} }}}} + 1  \\
   &=& \frac{3}{2}{\lambda _{\max }}\left( {{P_i}{\mb{T}}_i^H{{\mb{T}}_i} - {P_i}{\mb{H}}_i^H{{\mb{H}}_i},{\mb{I}} + {P_i}{\mb{H}}_i^H{{\mb{H}}_i}} \right) + 1 \hfill \\
   &\stackrel{(b)}\leq & {\left( {{\lambda _{\max }}\left( {{P_i}{\mb{T}}_i^H{{\mb{T}}_i} - {P_i}{\mb{H}}_i^H{{\mb{H}}_i},{\mb{I}} + {P_i}{\mb{H}}_i^H{{\mb{H}}_i}} \right) + 1} \right)^{\frac{3}{2}}}  \\
   &\stackrel{(c)}=& \left({\lambda _{\max }}\left( {{\mb{I}} + {P_i}{\mb{T}}_i^H{{\mb{T}}_i},{\mb{I}} + {P_i}{\mb{H}}_i^H{{\mb{H}}_i}} \right) \right)^{\frac{3}{2}},
    \end{eqnarray*}
where $(a)$ and $(c)$ are from $\lambda_{\max}(\mb{A},\mb{B}) = \lambda_{\max}(\mb{A}-\mb{B},\mb{B})+1$, $(b)$
is from $\frac{3}{2} x + 1 \leq x^{\frac{3}{2}}$ when $x$ is a nonnegative real number.

Therefore, we obtain $\mathcal{R}_{\max}^{DT} \geq \mathcal{R}_{\max}^{3P}$ when $P_R \to 0$. Together with $\mathcal{R}_{\max}^{2P}\rightarrow 0$ when $P_R \to 0$, we obtain Proposition \ref{corollary_P_R_0}.

\bibliographystyle{IEEEtran}
\bibliography{IEEEabrv,Two_way_secrecy_untrusted_Relay}

\end{document}